\documentclass[aps, twocolumn, showpacs, preprintnumbers, superscriptaddress, amsmath, amssymb, prb]{revtex4}
\usepackage{graphicx}
\usepackage{dcolumn}
\usepackage{bm}
\usepackage{color}

\begin{document}

\title{
Impurity-Induced Environmental Quantum Phase Transitions in the Quadratic-Coupling Spin-Boson Model}

\author{Da-Chuan Zheng}
\affiliation{Department of Physics, Renmin University of China, 100872 Beijing, China}
\author{Li Wan}
\affiliation{Department of Physics, Wenzhou University, 325035 Wenzhou, China}
\author{Ning-Hua Tong}
\email{nhtong@ruc.edu.cn}
\affiliation{Department of Physics, Renmin University of China, 100872 Beijing, China}
\date{\today}

\begin{abstract}
We study the zero temperature properties of the sub-Ohmic spin-boson model with quadratic spin-boson coupling. This model describes experimental set ups at the optimal working point where the linear-coupling between the qubit (spin) and the environmental noise (bosons) is zero and the leading coupling is quadratic. In the strong coupling regime, we find that the existence of spin induces quantum phase transitions (QPTs) between two states of environment: the normal state and a state with local distortions. The phase diagram contains both continuous and the first-order QPTs, with non-trivial critical properties obtained exactly. At the QPTs, the equilibrium state spin dynamics bears power-law $\omega$ dependence in the small frequency limit and a robust coherent Rabi oscillation at high frequency. We discuss the feasibility of observing such environmental QPTs in the qubit-related experiments.
\end{abstract}

\pacs{05.10.Cc, 64.70.Tg, 03.65.Yz, 05.30.Jp}


\maketitle

\section{Introduction}

The spin-boson model (SBM) is a frequently used paradigm to study the influence of the environmental noise on the quantum evolution of a two-level system.~\cite{Caldeira1,Leggett1,Weiss1}. The noise-induced dissipation and dephasing are the central issues of a variety of research fields, ranging from the electron/energy transfer in bio-chemical systems \cite{Garg1,Xu1,Song1,Muehlbacher1} to the endeavour of building a quantum computer.~\cite{Makhlin2,Shnirman1,Devoret1,You1,Novais1} Sufficiently strong coupling to the bosonic bath also induces a localized-delocalized quantum phase transition (QPT) in the two-level system.~\cite{Bulla1,Bulla1p,Florens1,Zhou1,Frenzel1,Vojta2} In recent years, there is much attention on the universality class of this QPT and debate on the applicability of the quantum-to-classical mapping in the deep sub-Ohmic regime of this system.~\cite{Vojta1,Vojta1p,Winter1,Alvermann1,Guo1,Glossop1,Kirchner1,Kirchner2} Experimental realization of the SBM have been proposed in various contexts, ranging from the mesoscopic metal ring to cold atom systems.~\cite{Tong1,Fedichev1,Recati1,Orth1,Porras1}

The SBM belongs to the impurity-bath problem for which the conventional focus is on the behavior of the impurity (a small quantum system) under the influence of the bath. 
For this purpose, the bath is regarded as stable and the influence from the impurity to bath is neglected in the conventional perturbative treatment. Such studies have been carried out for SBM in which a spin is coupled linearly to the displacement operator of 
harmonic oscillators. Note that recent study disclosed changes in the bath close to the QPT~\cite{Blunden-Codd1} in the linear-coupling SBM.

Recently, much attention is drawn to the SBM where a spin is coupled to the square of the boson displacement operator. It is triggered by the advances in the superconducting qubit experiments~\cite{Vion1,Bertet1,Ithier1,Yoshihara1,Kakuyanagi1} where the linear qubit-noise coupling is tuned to zero to suppress the decoherence, leaving the leading order coupling quadratic. The coherence time increases significantly at this optimal working point (OWP).~\cite{Bertet1} The quadratic coupling also appears in experiments based on semiconductor quantum dot~\cite{Petersson1} and the bismuth doners in silicon.~\cite{Wolfowicz1} Theoretically, the effect of quadratic coupling on the dephasing of qubit is analyzed.~\cite{Makhlin1,Bergli1,Cywinski1,Balian1} Quadratic electron-phonon coupling is used to explain the anomalous temperature dependence of the absorption line shape for quantum dot-based qubit systems.~\cite{Muljarov1,Borri1} Mohammad {\it et al.} suggested that the non-linear coupling leads to fundamentally different behaviour in the quantum Brownian motion.~\cite{Maghrebi1}

With a quadratic spin-boson coupling, the symmetry of the Hamiltonian is different from the linear coupling case, leading to different QPTs. 
Roughly speaking, in the linear case with positive coupling coefficient, positive $\langle S_z \rangle $ leads to $\langle X \rangle > 0$ (negative $\langle S_z \rangle$ to $\langle X \rangle < 0$) due to the $X S_z$ coupling, i.e., the symmetry breaking occurs in both $\langle S_z \rangle$ and $\langle X\rangle$ (here $X$ is the displacement operator of the environmental bosons). In this paper, we show that the strong quadratic coupling of the form $X^2 S_z$ induces positive or negative $\langle X \rangle$ if $\langle S_z \rangle < 0$, i.e., the symmetry breaking only occurs in $\langle X \rangle$. This QPT is thus regarded as an environmental QPT. 
 The ground state phase diagram contains first-order as well as continuous QPTs. Via such QPTs, the environment of the qubit changes from a normal state to a state with local distortions, leading to new dynamics both for the spin and the bath. These QPTs bear non-trivial critical properties amenable to experimental detection.

The rest of the paper is organized as follows. In Section II, we describe the model and the methods used to study it. Section III is devoted to the main results, including the exact solution at $\Delta=0$ and the NRG results for $\Delta > 0$. Various related issues of the quadratic-coupling SBM are discussed in Section IV. The details of the exact solution at $\Delta=0$ is presented in Appendix A. The NRG formalism is summarized in Appendix B. In Appendix C, we present quantitative comparison between NRG data and the exact solution at $\Delta=0$.

\section{Model and Methods}

A general Hamiltonian describing a two-level system coupled to environmental noise can be written as

\begin{equation}
  H = \frac{\epsilon}{2} \sigma_z -\frac{\Delta}{2} \sigma_x + \displaystyle\sum_{i} \omega_{i} a_{i}^{\dagger}a_{i} + \frac{1}{2} \sigma_{z} f(\hat{Y}) ,
\end{equation} \label{eq1}
where $\hat{Y} = \sum_{i} \lambda_{i}(a_{i}+a_{i}^{\dagger})$ is the local boson displacement operator. $\lambda_i$ describes the local weight of the $i$-th boson mode. The two-level system is represented by a spin $1/2$ with bias $\epsilon$ and tunnelling strength $\Delta$. It is coupled to the bosonic bath with mode energies $\{ \omega_i \}$ in terms of $\sigma_z$ and $\hat{Y}$. 
In the weak coupling limit, the function $f(z)$ can be expanded into Taylor series $f(z)=g_{0} + g_{1}z + g_{2}z^{2} +... $. The conventional SBM Hamiltonian is obtained by truncating the series at the linear order. At the OWP of the superconducting qubit experiments \cite{Vion1,Bertet1,Ithier1,Yoshihara1,Kakuyanagi1,
Petersson1,Wolfowicz1} and in other experimental setups,~\cite{Petersson1,Wolfowicz1} $g_{1}$ is zero and the leading coupling is quadratic in the boson coordination.~\cite{Makhlin1} Truncating the series at this order and absorbing the constant $g_{0}$ into $\epsilon$, we obtain the Hamiltonian of the quadratic-coupling SBM,
\begin{equation}
  H_{QSB} = \frac{\epsilon}{2} \sigma_z -\frac{\Delta}{2} \sigma_x + \displaystyle\sum_{i} \omega_{i} a_{i}^{\dagger}a_{i} + \frac{g_2}{2}  \sigma_{z} \hat{Y}^{2}.
\end{equation}  \label{eq2}
The effect of the bath on the spin is encoded into the bath spectral function $J(\omega)$ defined as
\begin{equation}
   J(\omega) =\pi \sum_{i} \lambda_{i}^{2} \delta(\omega -
   \omega_{i}).
\end{equation}  \label{eq3}
In this paper we mainly focus on the continuous bath with a power law spectrum in small $\omega$ limit and a hard-cutoff at $\omega = \omega_c$,
\begin{equation}
   J(\omega) = 2 \pi \alpha \omega^{s} \omega_c^{1-s}   \,\,\,\,\,\,(0 < \omega < \omega_c) ,  \label{eq4}
\end{equation}
which includes the most frequently encountered cases in experiments such as the $1/f$ noise.~\cite{Makhlin1} Quantitative prediction for the single-mode Hamiltonian of the qubit-resonator experiment~\cite{Bertet1} will be discussed in the end of this paper.
The coupling constant $g_2$ can be absorbed into $\lambda_i$, or equivalently, is set as unity in the numerical calculation below. 
In Eq.(4), $\alpha$ controls the strength of the spin-boson coupling. Our study is confined to the sub-Ohmic bath with $0 < s < 1$ and the conclusion is extended afterwards to the Ohmic case $s=1$ and to $s=0$ for the $1/f$ noise. $\omega_c=1.0$ is set as the energy unit. 

Here we compare the symmetry of $H_{QSB}$ to that of the linear-coupling SBM $H_{LSB}$. At $\epsilon=0$, $H_{LSB}$ is invariant under the combined boson and spin transformation $U a_{i} U^{-1} = -a_i$ and $U \sigma_{z} U^{-1} = - \sigma_z$. Previous studies disclosed that for the sub-Ohmic ($0 \leqslant s<1$) and the Ohmic ($s=1$) baths, a strong coupling strength may induce a spontaneous breaking of this symmetry and the system enters the localized phase, in which the quantum system is trapped to one of the two states and the local bosons have finite displacements.~\cite{Bulla1,Bulla1p} This transition is the so-called delocalized-localized transition of the SBM. 

With a quadratic coupling, $H_{QSB}$ is invariant under the parity transformation $U a_{i} U^{-1} = -a_i$ alone. In case the spin is in the state $\sigma_z < 0$, the quadratic coupling contributes negative energies for boson modes proportional due to $- \langle \hat{Y}^2 \rangle$. When overcoming the positive energies $\omega_i$ of the low energy boson modes, They lead to an instability of the bosons. Physically, as the coupling strength increases, the harmonic potentials of the environmental particles are softened and the instability occurs when the potential wells are inverted, accompanied with the divergence of particle numbers. At this transition, the boson parity symmetry is spontaneous broken. Taking into account the boson anharmonic potentials that are neglected in $H_{QSB}$, this instability will lead to a local distortion in the environmental degrees of freedom. Even for a weak quadratic coupling strength, the feedback effect of the impurity to the bath can no longer be regarded as small and the bath is intrinsically non-Gaussian. New dynamical behaviour will emerge both in the bath and in the impurity.

Such QPTs can be studied exactly at the non-tunnelling point $\Delta=0$ at which $[\sigma_z, H_{QSB} ] =0$. The eigen-states of $H_{QSB}$ are in the form $| \Psi^{(+1)} \rangle |+1 \rangle$ and $| \Psi^{(-1)} \rangle |-1 \rangle$. $|+1 \rangle$ and $|-1\rangle$ are eigen-states of $\sigma_z$ with energies $+1$ and $-1$, respectively. $| \Psi^{(\pm 1)} \rangle$ are the corresponding boson states. In each spin sector, the quadratic boson Hamiltonian can be solved exactly. We use the equation of motion method for the double-time Green's functions to obtain the exact properties of $H_{QSB}$ at $\Delta=0$. The derivation is summarized in Appendix A.

For general parameters, we study $H_{QSB}$ using the Wilson's numerical renormalization group (NRG) method \cite{Wilson1, Bulla2} adapted to bosonic bath.~\cite{Bulla1,Bulla1p} The Wilson chain Hamiltonian can be derived from an orthogonal transformation of the logarithmic-discretized bath. It is given as
\begin{eqnarray}
   H_{NRG}  &=& \frac{\epsilon}{2} \sigma_z -\frac{\Delta}{2} \sigma_x + \displaystyle\sum_{n=0}^{\infty} \left[ \epsilon_{n} b_{n}^{\dagger}b_{n} + t_{n}\left(b_{n}^{\dagger}b_{n+1} + b_{n+1}^{\dagger}b_{n} \right) \right]  \nonumber \\
   && + \frac{g_2}{2} \frac{\eta_0}{\pi} \sigma_{z} \hat{X}^{2}.
\end{eqnarray}   \label{eq5}
Here $\epsilon_n, t_n \propto \Lambda^{-n}$ are the on-site and hopping energies of the boson chain and $\Lambda \geqslant 1$ is the logarithmic discretization parameter. The displacement operator $\hat{Y}$ in Eq.(2) is normalized as $\hat{Y} = \sqrt{\eta_0/ \pi} \hat{X}$ with $\hat{X} = b_0 + b_0^{\dagger}$. The local boson annihilation operator reads
\begin{equation}
b_{0} = \sqrt{ \frac{\pi}{\eta_0}} \sum_i \lambda_i a_i.
\end{equation}   \label{eq6}
Here $\eta_0 = \pi \sum_i \lambda_i^{2} = \int_{0}^{\infty} J(\omega) d\omega$. The formalism used for NRG calculation is summarized in Appendix B.
Thanks to the exponential decay of energy scales along the chain, the low energy eigen-energies and eigen-states of $H_{NRG}$ can be obtained reliably by iteratively diagonalizing $H_{QSB}$ and keeping the lowest $M_s$ states after each diagonalization. For each boson site to be added into the chain, we truncate its infinite dimensional Hilbert space into a $N_b$-dimensional space on the occupation number basis.
The accuracy of NRG result is controlled by three parameters: the logarithmic discretization parameter $\Lambda$, the number of kept states $M_{s}$, and the boson-state truncation parameter $N_{b}$. In this work, we obtain the exact results at $\Lambda = 1.0$, $M_s = \infty$, and $N_{b} = \infty$ by extrapolating the NRG data from $\Lambda = 1.6 \sim 10.0$, $Ms = 60 \sim 300$, and $N_{b}= 8 \sim 50$ to the above limit.

Here a remark on the applicability of the NRG is in order. Previous studies of the QPT in linear-coupling SBM showed that naive application of NRG gives incorrect exponents $\beta$, $\delta$, and $x$ in the deep sub-Ohmic regime ($0 \leqslant s < 1/2$),~\cite{Vojta1} due to the boson state truncation error~\cite{Hou1,Tong2} and the mass flow error.~\cite{Vojta1p} These errors only influence the order parameter related exponents $\beta$ and $\delta$, and the susceptibility-temperature exponent $x$ defined at the critical point. In our NRG study below, we study the critical behavior from the weak-coupling side of the QPT and avoid those possibly problematic exponents. We check the $N_b$ dependence of the critical behavior to exclude the possibility of boson state truncation error. We also compare the NRG results with the exact solution at $\Delta=0$. The perfect agreement in the exponents strongly supports the reliability of our NRG calculation.

\section{Results}

$H_{QSB}(\Delta=0)$ contains all the non-trivial properties of the environmental QPTs excepts for the dynamics of $\sigma_z$. A finite quantum tunnelling $\Delta > 0$ introduces non-trivial dynamics of $\sigma_z$ but only modifies the phase diagram quantitatively. 
Below, we first study the $\Delta=0$ case, presenting the exact solution as well as the NRG results. Then, we use NRG to study the effect of finite quantum tunnelling $\Delta > 0$.

\subsection{Non-tunnelling point $\Delta = 0$}

The Hamiltonian $H_{QSB}$ at $\Delta=0$ reads
\begin{equation}
    H_{QSB}(\Delta=0) =  \frac{\epsilon}{2} \sigma_z + \displaystyle\sum_{i} \omega_{i} a_{i}^{\dagger}a_{i} + \frac{g_2 \eta_0}{2\pi}  \sigma_{z} \hat{X}^{2}.
\end{equation}   \label{eq7}
Here $\hat{X}$ is the normalized boson displacement operator defined in Eq.(5). At this exact soluble limit, the dephasing properties were analysed in the context of the superconducting qubit at the optimal working point \cite{Makhlin1} and the quantum dot qubit quadratically coupled to acoustic phonons \cite{Muljarov1}. As confirmed by our NRG calculation below, the universal critical properties of the QPTs for general $H_{QSB}(\Delta)$ are already well described by this limit. 

\subsubsection{exact solution for $\Delta=0$}
The change of the environment by the presence of impurity is best seen in the effective boson spectral function 
\begin{equation}
   C_{X}(\omega) =\frac{1}{2\pi} \int_{-\infty}^{+\infty} C_{X}(t) e^{i\omega t} \, dt  ,
\end{equation}   \label{eq8}
with $C_{X}(t) \equiv (1/2) \langle \{ X(t), X(0) \} \rangle$. We calculate the exact expression for $C_{X}(\omega)$ and the ground state energy difference $\Delta E_{g} \equiv E_{g}^{(+1)} - E_{g}^{(-1)}$ between the two subspaces $\sigma_z = \pm 1$, from which the exact ground state phase diagram can be extracted.
Using the Green's function equation of motion method, the exact expression for $C_{X}(\omega)$ at $T=0$ is obtained as (see Appendix A for details),
\begin{equation}
    C_{X}(\omega) = \frac{\frac{1}{2\eta_0}J(\omega) }{ \left\{ 1 - g_{2} \frac{\eta_0}{\pi} \sigma_z \left[g(\omega) + g(-\omega) \right] \right\}^2 + g_{2}^{2} J^{2}(\omega)}
\end{equation}   \label{eq9}
for $\omega > 0$. For $\omega < 0$, $C_{X}(\omega) = C_{X}(-\omega)$. 
The function $g(\omega)$ is given as
\begin{equation}
   g(\omega) = \frac{1}{\eta_0} \mathcal{P}\int_{0}^{\infty} \frac{J(\epsilon)}{\omega - \epsilon} d\epsilon.
\end{equation}   \label{eq10}
For the specific $J(\omega)$ in Eq.(4), $\eta_0= 2\pi\alpha \omega_c^{2}/(1+s)$ and 
\begin{equation}
   g(\omega) = \frac{1}{\omega} F(1, 1+s; 2+s; \frac{\omega_c}{\omega}).
\end{equation}   \label{eq11}
Here $F(\alpha, \beta; \gamma; z)$ is the hypergeometric function.

In the weak coupling limit $\alpha=0$, $C_{X}(\omega) = J(\omega)/(2\eta_0)$ recovers the normalized bare spectral function.
A finite quadratic coupling to the impurity exerts significant influence on $C_{X}(\omega)$. In particular, in the subspace $\sigma_z = -1$, a singularity develops in $C_{X}(\omega=0)$ at $\alpha = \alpha_c$ which signals a continuous QPT. Using the analytical continuation of $F(\alpha, \beta; \gamma; z)$ from $|z| > 1$ to $|z|<1$, and considering $F(\alpha, \beta; \gamma; z=0)=1$, we find $\alpha_c = s/(4 g_2 \omega_c)$. No QPT occurs for $\alpha > 0$ in the other subspace $\sigma_z = +1$. We denote the asymptotic behaviour of $C_{X}(\omega)$ in the small frequency limit as $C_{X}(\omega) \propto \omega^{y_0}$ for $\alpha < \alpha_c$ and $C_{X}(\omega) \propto \omega^{y_c}$ at $\alpha = \alpha_c$. The exact solution reads
\begin{eqnarray}
  C_X(\omega) = \left\{
\begin{array}{lll}
\frac{2(1+s)}{\pi^{2}s^{2}}\frac{1}{\omega_c} \left( \frac{\omega}{\omega_c} \right)^{-s}, \,\,\,\,\,\,&  (\alpha = \alpha_c);\\
& \\
\frac{(1+s)\alpha_c^{2}}{2\left( \alpha_c - \alpha \right)^2} \frac{1}{\omega_c} \left( \frac{\omega}{\omega_c} \right)^s ,\,\,\,\,\,\,
& (\alpha < \alpha_c).
\end{array} \right.
\end{eqnarray}   \label{eq12}
This gives $y_0 = s$ and $y_c = -s$.  
For a fixed $\alpha < \alpha_c$, $C_{X}(\omega) \sim (\omega/\omega_c)^{s}$ for $\omega \ll \omega^{\ast}$ and $C_{X}(\omega) \sim (\omega/ \omega_c)^{-s}$ for $\omega \gg \omega^{\ast}$. There is a peak at the crossover frequency $\omega = \omega^{\ast}$, with
\begin{equation}
\omega^{\ast} = \omega_c \left(\frac{4}{\pi^2 s^2 \alpha_c^2} \right)^{\frac{1}{2s}} (\alpha_c - \alpha)^{\frac{1}{s}},  \,\,\,\,\,\,\,\, (\alpha < \alpha_c).
\end{equation}   \label{eq13}
It corresponds to the crossover energy scale $T^{\ast}$ between the boson-stable state and the quantum critical regime.
As $\alpha$ approaches $\alpha_c$ from below, the peak position moves to zero frequency in a power law $\omega^{\ast} \propto (\alpha_c - \alpha)^{z\nu}$, giving the exact exponent $z\nu = 1/s$.

The two subspaces $\sigma_z = \pm 1$ have the ground state energy difference
\begin{equation}
   \Delta E_{g} =  E_{g}^{(+1)} - E_{g}^{(-1)} = \epsilon + \frac{1}{\pi} \int_{-\infty}^{0} \text{Im} H(\omega + i \eta) d\omega .
\end{equation}   \label{eq14}
Here, $\eta$ is an infinitesimal positive number and
\begin{equation}
    H(\omega) = \frac{2g_2 / \pi}{1-(2g_2 / \pi)^{2} h^{2}(\omega)} \left[ h(\omega) + k(\omega) \right] ,
\end{equation}   \label{eq15}
with
\begin{eqnarray}
&&  h(\omega) = \int_{-\infty}^{\infty} \frac{J(\epsilon) \epsilon}{\omega^2 - \epsilon^2} d\epsilon,  \nonumber  \\
&&  k(\omega) = \int_{-\infty}^{\infty} \frac{J(\epsilon) \epsilon}{ (\omega - \epsilon)^2} d\epsilon . 
\end{eqnarray}   \label{eq16}
For a fixed coupling strength $\alpha$, $E_g^{(+1)}< E_g^{(-1)}$ for very large negative $\epsilon$. $E_g^{(+1)}$ increases with increasing $\epsilon$. A level crossing occurs at $\epsilon = \epsilon_f$, at which the global ground state change from the subspace $\sigma_z=1$ to $\sigma_z=-1$. The spin-flip transition point $\epsilon_f(\alpha)$ is determined by $\Delta E_{g}(\epsilon_f) =0$. Taylor expanding $\Delta E_{g}$ with respect to $\alpha$ and solving this equation, we obtain in the small $\alpha$ limit
\begin{equation}
   \epsilon_f = - \frac{2 \alpha}{1+s} (g_2 \omega_c^2) + \mathcal{O}(\alpha^3).
\end{equation}   \label{eq17}

\subsubsection{NRG results for $\Delta=0$}
\begin{figure}[t!]
\vspace{-1.8cm}
\begin{center}
\includegraphics[width=3.9in, height=3.0in, angle=0]{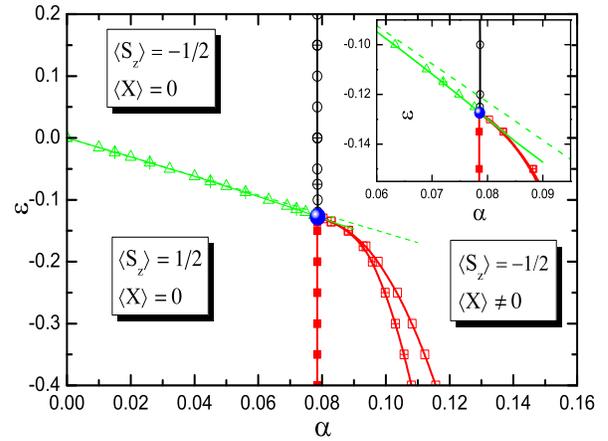}
\vspace*{-1.0cm}
\end{center}
\caption{(Color online) Ground state NRG phase diagram of $H_{QSB}(\Delta=0)$ for $s=0.3$. $\langle X \rangle = \langle b_{0}^{\dagger}+ b_{0} \rangle$ is the order parameter and $\langle S_{z}\rangle = \langle \sigma_z \rangle/2$ is the spin polarization. The phase boundaries are the spin flip (up triangles), continuous QPT (circles), and first-order QPT (squares) lines. Lines are for guiding eyes. NRG parameters are $N_{b}=8$ (empty symbols), $N_{b}=12$ (plus-filled symbols), and the extrapolated $N_{b}=\infty$ (solid squares). $(\alpha_{c}=0.0786, \epsilon_{c}= -0.1273)$ (soild dot) is the jointing point of the three transition lines). Inset: details close to the jointing point, with the exact spin flip line (solid line) and the weak-coupling expansion $\epsilon_f = -2 \alpha / (1+s)$ (dashed line). NRG parameters are $\Lambda=2.0$ and $M_{s}=60$.}   \label{Fig1}
\end{figure}

We further study the nature of QPTs at $\Delta=0$ using bosonic NRG. Quantitative comparison of NRG results with the exact solution (see Appendix C) shows perfect agreement, which benchmarks our NRG calculation. For simplicity, we present results only for a generic sub-Ohmic bath  $s=0.3$. Unless specified otherwise, qualitatively similar results are obtained for other $s$ values. 

Fig.1 shows the ground state phase diagram on the $\alpha-\epsilon$ plane. Phases are characterized by different values of the spin polarization $\langle S_z \rangle$ and the order parameter $\langle X \rangle$. The phases with $|\langle X \rangle| =0$ and $|\langle X \rangle| \neq 0$ are called environment-stable and -unstable phases, respectively. The boson parity symmetry is spontaneously broken in the latter. Phase boundaries are obtained using NRG with $N_b=8$ and $12$. The spin flip at $\epsilon=\epsilon_f(\alpha)$ and the continuous QPT at $\alpha=\alpha_c^{(c)}$ are found insensitive to $N_b$, while the first-order QPT line at $\alpha = \alpha_c^{(1)}$ moves with $N_b$ and converges in the limit $N_b= \infty$ to the same vertical line as the continuous QPT (solid squares with guiding line), giving $\alpha_c^{(1)}=\alpha_c^{(c)} = 0.0786$. This value is slightly larger than $\alpha_c^{{\text exc}} = s/(4g_2 \omega_c) = 0.075$, due to the logarithmic discretization error at $\Lambda=2.0$. Extrapolating $\alpha_c^{(c)}$ to $\Lambda=1.0$ gives perfect agreement with $\alpha_c^{{\text exc} }$, as shown in Fig.C2 of Appendix C. The three transition lines meet at a jointing point ($\alpha_c=0.0786$, $\epsilon_c=-0.1273$) (solid dot in Fig.1).

The inset of Fig.1 shows details close to the jointing point. There is very good agreement in the spin-flip line $\epsilon_f$ from NRG using $N_b=8$ and $\Lambda=2.0$ (up triangles) and the exact solution from $\Delta E_g =0$ (solid line). This is due to the cancellation of errors of $E_g^{(+1)}$ and $E_{g}^{(-1)}$ in the NRG calculation, since the error in the NRG ground state energy comes from its treatment of bosons, independent of the spin state.

The first-order QPT is a level crossing induced by the boson instability transition in the subspace $\sigma_z = -1$. For $\epsilon < \epsilon_c$ and small $\alpha$, the subspace $\sigma_z=-1$ has higher energy than the $\sigma_z=1$ subspace. As we increases $\alpha$ to $\alpha = \alpha_c^{(c)}$, $E_g^{(-1)}$ decreases abruptly to $-\infty$ at the boson-unstable QPT in this subspace, leading to a sharp crossing of $E_g^{(-1)}$ and $E_g^{(+1)}$. This scenario of the QPTs suggests $\alpha_{c}^{(1)} = \alpha_{c}^{(c)}$ for $\Delta=0$, both being independent of $\epsilon$. Indeed, although NRG gives an $\epsilon$-dependent $\alpha_c^{(1)}$ for finite $N_{b}$, as shown in Fig.C3 of Appendix C, it converges to the vertical line at $\alpha = \alpha_c^{(c)}$ in the limit $N_b \rightarrow \infty$ (solid squares in Fig.1).

\begin{figure}[t!]
\vspace{-4.0cm}
\begin{center}
\includegraphics[width=5.6in, height=4.1in, angle=0]{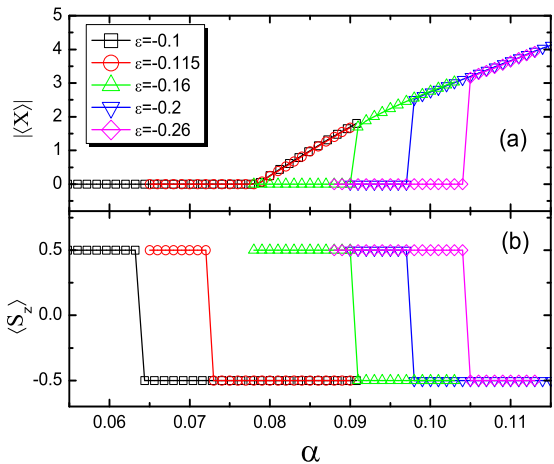}
\vspace*{-1.0cm}
\end{center}
\caption{(Color online) NRG results for $|\langle X \rangle|$ and $\langle S_z \rangle$ as functions of $\alpha$ for $s=0.3$ and $\Delta=0.0$, for various $\epsilon$ values. The NRG parameters are $\Lambda=2.0$, $M_{s}=60$ and $N_{b}=8$. 
}     \label{Fig2}
\end{figure}

In Fig.2, $\langle X \rangle$ and $\langle S_z \rangle$ are plotted as functions of $\alpha$ for various $\epsilon$ values. For $\epsilon=-0.1$ and $-0.115$ which are larger than $\epsilon_c$, as $\alpha$ increases, a spin-flip transition occurs first (jumps in Fig.2(b)) and it is followed by a continuous QPT at larger $\alpha$ (continuous emerging of nonzero $|\langle X \rangle |$ at $\alpha = \alpha_c^{(c)} $ in Fig.2(a)). For $\epsilon=-0.16$, $-0.2$, and $-0.26$ which are smaller than $\epsilon_c$, both quantities jump discontinuously at $\alpha_c^{(1)}$. The phase diagram can be mapped out from such plots. It is noted that Fig.2(a) shows only the qualitative behavior of $\langle X \rangle$ for finite $N_{b}$. In the limit $N_b = \infty$, as shown in Fig.C1 and Fig.C3 in Appendix C, $|\langle X \rangle|$ diverges both at the continuous and the first-order QPTs, being consistent with the scenario that the harmonic potentials of the bath oscillators are inverted at $\alpha > \alpha_c^{(c)}$ or $\alpha > \alpha_c^{(1)}$.

\begin{figure}[t!]
\vspace{-2.0cm}
\begin{center}
\includegraphics[width=4.1in, height=3.5in, angle=0]{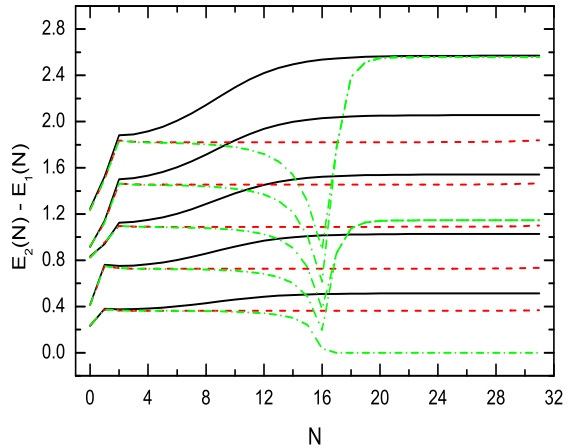}
\vspace*{-0.8cm}
\end{center}
\caption{(Color online) NRG flow of excitation energies at $s=0.3$, $\Delta=0.0$ and $\epsilon=0.1 > \epsilon_c$. The energy levels flow to three different fixed points: a stable free boson fixed point for $\alpha=0.084 < \alpha_c^{(c)}$ (solid lines), an unstable critical fixed point for $\alpha=0.08593945 \approx \alpha_c^{(c)}$ (dashed lines), and a strong-coupling fixed point for $\alpha=0.086 > \alpha_c^{(c)}$ (dash-dotted lines). The NRG parameters are $\Lambda=4.0$, $M_s=100$, and $N_{b}=8$.}   \label{Fig3}
\end{figure}

%
\begin{figure}[t!]
\vspace*{-2.0cm}
\begin{center}
\includegraphics[width=4.6in, height=3.4in, angle=0]{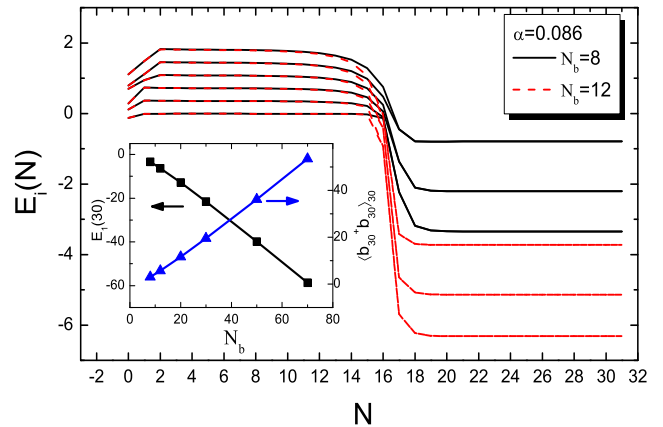}
\vspace*{-1.0cm}
\end{center}
\caption{(Color online) NRG flow of $E_{i}(N)$ ($i=1,2,3,...,6$) at $s =0.3$, $\Delta=0.0$, $\epsilon=0.1 > \epsilon_c$, and $\alpha=0.086 > \alpha_c^{(c)}$, obtained using $N_b=8$ (solid lines) and $N_b=12$ (dashed lines). Inset: ground state properties of the fixed point Hamiltonian $H_{N=30}$ as functions of $N_b$, the energy $E_{1}(N=30)$ (squares) and the boson occupancy number $\langle b_{30}^{\dagger}b_{30} \rangle_{N=30}$ (up triangles). The lines are for guiding eyes. Other NRG parameters are $\Lambda=4.0$ and $M_s=100$.}   \label{Fig4}
\end{figure}

%
\begin{figure}[t!]
\begin{center}
\includegraphics[width=5.5in, height=4.5in, angle=0]{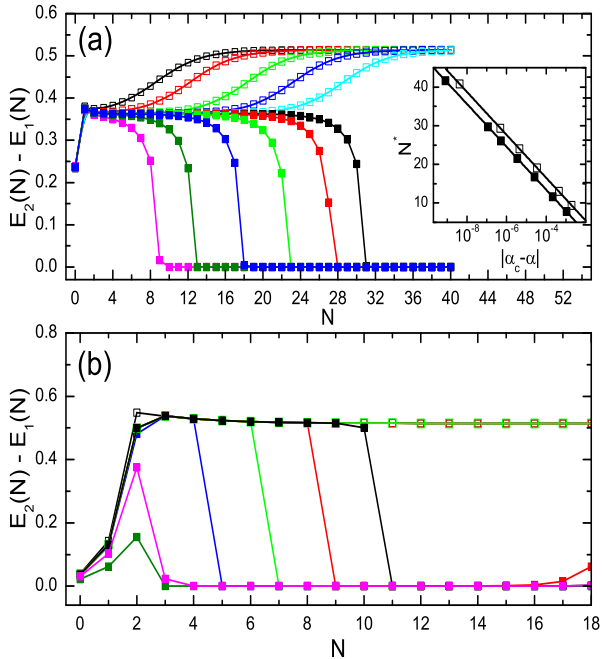}
\vspace*{-2.4cm}
\end{center}
\caption{(Color online) Flow diagrams near the QPTs for $s=0.3$ and $\Delta=0.0$. (a) continuous QPT for $\epsilon=0.1$, $\alpha_c \approx 0.085934$, and (b) first-order QPT for $\epsilon=-0.2$, $\alpha_c \approx 0.10147$. From left to right, $\alpha$ increases for $\alpha < \alpha_c$ (empty squares) and $\alpha$ decreases for $\alpha > \alpha_c$ (solid squares). Lines are for guiding eyes. Inset of (a): power law fitting of $T^{\star} = \Lambda^{-N^{\star}} \propto |\alpha - \alpha_{c}|^{z\nu}$ for $\alpha < \alpha_c$ (empty squares) and $\alpha > \alpha_c$ (solid squares), giving $z\nu = 3.333$ and $3.338$, respectively. NRG parameters are $\Lambda=4.0$,  $M_s=100$ and $N_{b}=12$.
}   \label{Fig5}
\end{figure}

As a direct product of NRG, the flow of the energy levels can help identify various fixed points in the parameter space. These fixed points are reflected by the mass of boson excitations in the exact bosonic Green's function. 
As shown in Fig.3, we found three distinct fixed points for $\epsilon=0.1 > \epsilon_c$. The stable fixed point obtained for $\alpha = 0.084 < \alpha_c^{(c)}$ is identified as the free boson fixed point with $\langle X \rangle=0$ and $\langle S_z \rangle =-1/2$. For $\alpha = 0.086 > \alpha_c^{(c)}$, the excitation energies flow towards a state with two-fold degeneracy. At this fixed point, the harmonic potential of bath particles is inverted and $\hat{X}$ fluctuates between $\pm \infty$. In the large $N$ regime, the numerical error will lift the degeneracy and break the boson parity symmetry, giving $\langle X \rangle \neq 0$.~\cite{Note1} At $\alpha = \alpha_c^{(c)}$, the excitation energies flow to an unstable critical fixed point and $\langle X \rangle$ begins to be nonzero continuously at this point.

To study the nature of the ordered phase, we plot in Fig.4 the flow of eigen-energies $E_{i}(N)$ ($i=1 \sim 6$), directly obtained from diagonalizing the Wilson chain Hamiltonian $H_N$, without subtracting the ground state energy $E_{1}(N)$. This is done for $\epsilon > \epsilon_c$ and $\alpha$ slightly larger than $\alpha_c^{(c)}$, {\it i.e.}, in the boson-unstable phase. The energies for $N < 14$ is independent of $N_b$ since the flow is still in the weak-coupling regime. The strong-coupling fixed point is reached for $N > 16$ and in that regime, the energies decrease with increasing $N_b$. Note that the excitation energies, {\it i.e.}, the differences between the energy levels, do not change significantly with $N_b$, including the two-fold degeneracies. As shown in the inset, both $E_{1}(N=30)$ and the boson number $\langle b^{\dagger}_{30} b_{30} \rangle_{N=30}$ at the strong-coupling fixed point $N=30$ are linear functions of $N_b$, diverging in the limit $N_b =\infty$. As a result, the total NRG ground state energy $E_{QSB} =  \sum_{n=0}^{\infty}  \Lambda^{-n} E_{1}(n)$ tends to negative infinity in the limit $N_b = \infty$. This supports that the strong-coupling fixed point is the environment-unstable state with inverted harmonic potentials for the bosonic modes.

To investigate the critical behaviour of the QPTs, the excitation energy flows are presented in Fig.5 for $\alpha$ very close to $\alpha_c^{(c)}$ and $\alpha_c^{(1)}$. In Fig.5(a), a typical critical behaviour is observed for $\epsilon=0.1 > \epsilon_c$, with the standard scaling form. The crossover energy scale $T^{\ast}  = \Lambda^{-N^{\ast}}$ is found to follow a power law, $T^{\ast} \propto |\alpha_c - \alpha|^{z\nu}$. The fitted exponent $z\nu=3.333$ and $z\nu=3.338$ from the two sides of $\alpha_c^{(c)}$ agree well with the exact solution $z\nu=1/s$ at $s=0.3$. In Fig.5(b), near the first-order phase transition at $\epsilon=-0.2 < \epsilon_c$, a level crossing in the energy flow is observed, accompanied with an abrupt jump from $S_z=1/2$ to $S_z=-1/2$ .

\subsection{Effects of finite quantum tunnelling $\Delta > 0$}

\subsubsection{ the case of s=0.3 }
\begin{figure}[t!]
\vspace*{-4.0cm}
\begin{center}
\includegraphics[width=5.7in, height=4.0in, angle=0]{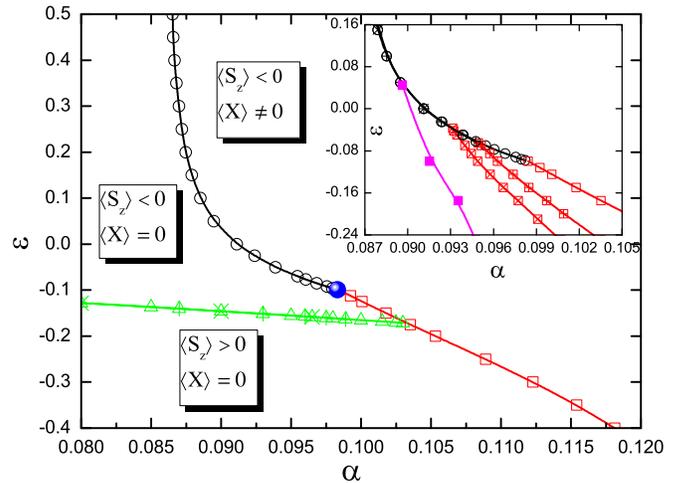}
\vspace*{-1.0cm}
\end{center}
\caption{(Color online) Main figure: ground state phase diagram for $s=0.3$ and $\Delta=0.1$, obtained with $N_b=12$. The continuous QPT, first-order QPT, and spin flip lines are marked by empty circles, empty squares, and empty up triangles with eye-guiding lines, respectively. The jointing point of the continuous and the first-order QPTs is marked by a solid blue dot.
The spin flip line obtained with $N_b =20$ (plus-filled up triangles) and $N_b=30$ (cross-filled up triangles) are also plotted.
Inset: the change of phase boundaries with $N_b$. $N_b=12$ (empty symbols), $N_b=20$ (plus-filled symbols), $N_b=30$ (cross-filled symbols), and the extrapolated $N_b = \infty$ (solid squares). NRG parameters are $\Lambda=4.0$ and $M_s=100$.   
}   \label{Fig6}
\end{figure}

The quantum tunnelling $\Delta > 0$ introduces non-trivial dynamics for $\sigma_z$ but only modifies the phase diagram quantitatively.

Fig.6 shows the NRG phase diagram for $s=0.3$ and $\Delta=0.1$. The boson-stable state ($\langle X \rangle=0$) on the left side is separated from the boson-unstable phase ($\langle X \rangle \neq 0$) on the right by a continuous (for $\epsilon > \epsilon_c)$ or a first-order (for $\epsilon < \epsilon_c$ ) QPT. The two QPT lines meet at the jointing point $(\alpha_c, \epsilon_c)$ (blue dot). The QPT lines in the main figure are obtained with $N_b=12$. The spin flip lines are obtained with $N_b=12$, $20$, and $30$ and they fall onto the same line, showing that the spin-flip line is independent of $N_b$, same as the $\Delta=0$ case. In order to show the $N_b$ dependence of the phase diagram, in the inset, we show the QPT lines for various $N_b$ values. It is seen that the continuous QPT line $\alpha_c^{(c)}$ (circles) is almost independent of $N_b$, while the first order QPT line $\alpha_c^{(1)}$ (squares) decreases with increasing $N_b$, converging to an extrapolated line in the limit $N_b= \infty$ (solid squares).  

The finite quantum tunnelling induces several changes with respect to $\Delta=0$. First, the phase boundaries shift quantitatively. The continuous QPT line is no longer vertical but depends on $\epsilon$, especially near $\epsilon_c$ where the competition between $\epsilon$ and $\Delta$ is strong. For $\epsilon \gg \Delta$, $\alpha_c^{(c)}$ is independent of $\epsilon$ asymptotically. The jointing point of the continuous and first-order QPT lines shifts upwards. 
Second, Due to the mixing of $\sigma_z=\pm 1$ subspaces by $\Delta > 0$, physical quantities change smoothly at the spin-flip line $\epsilon_f(\alpha)$ which only marks $\langle S_z \rangle =0$ and has no quantum fluctuations. 
The ending point of the spin-flip line lies on the first-order line and is below the jointing point of continuous and first-order QPT lines. 
Third, for $\Delta=0$, the QPT from a $\langle S_z\rangle < 0$ state to another $\langle S_z\rangle < 0$ state by increasing $\alpha$ is always continuous. 
In contrast, for $\Delta > 0$, a small $\epsilon$ window (below the jointing point and above the ending point of the spin-flip line) opens, in which the QPT from a $\langle S_z\rangle < 0$ state to another $\langle S_z\rangle < 0$ state is first order. 
For this $\epsilon$ regime, although $\langle Sz \rangle <0$ on the $\alpha < \alpha_c$ side, due to spin fluctuations, the ground state contains a finite components of spin up states. When $\alpha$ increases, according to the scenario built at $\Delta=0$, the spin up components tend to change into spin down state abruptly, making the transition first order.

\begin{figure}[t!]
\vspace*{-5.0cm}
\begin{center}
\includegraphics[width=6.0in, height=5.2in, angle=0]{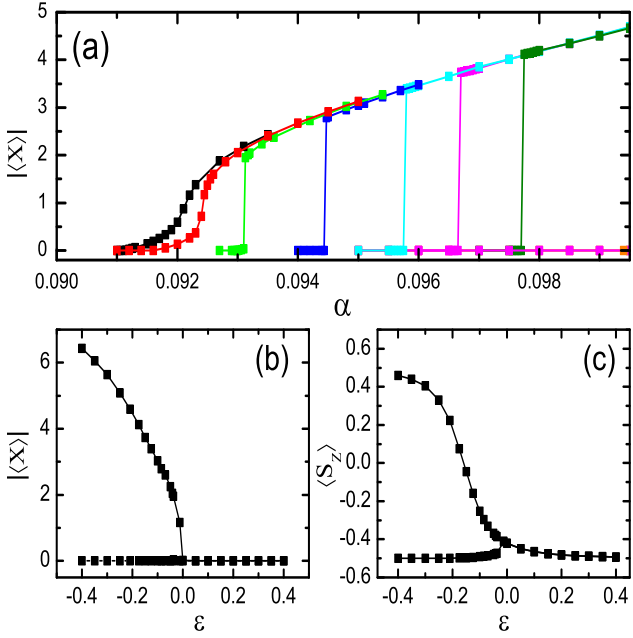}
\vspace*{-1.0cm}
\end{center}
\caption{(Color online) $|\langle X \rangle|$ and $\langle S_z \rangle$ near the jointing point of the continuous and the first-order QPTs, for $s=0.3$ and $\Delta=0.1$. (a) $|\langle X \rangle|(\alpha)$ for various $\epsilon$'s. From right to left, $\epsilon= -0.175$, $-0.15$, $-0.125$, $-0.085$, $-0.037$, $-0.012$, and $0.0$. In (b) and (c),  $|\langle X \rangle|$ and $\langle S_z \rangle|$ values at the upper and lower edge of the transition as functions of $\epsilon$. NRG parameters are $\Lambda=4.0$, $M_s=100$, $N_b=30$.
}   \label{Fig7}
\end{figure}

%
\begin{figure}[t!]
\vspace*{-4.5cm}
\begin{center}
\includegraphics[width=5.7in, height=4.8in, angle=0]{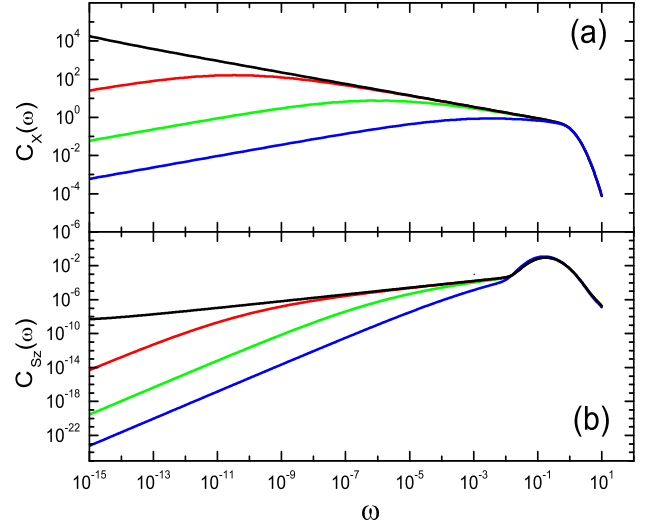}
\vspace*{-1.0cm}
\end{center}
\caption{(Color online) Dynamical correlations (a) $C_{X}(\omega)$ and (b) $C_{Sz}(\omega)$ for $\alpha \leq \alpha_{c}^{(c)}$ for $s=0.3$, $\Delta=0.1$, and $\epsilon=0.0 > \epsilon_{c}$. From top to bottom, $\alpha = 0.091101 \approx \alpha_c^{(2)}$, $0.08$, $0.09$, and $0.09105$. The fitted exponents are for (a): $y_0 = 0.296$ for $\alpha < \alpha_c$, $y_c = -0.302$ for $\alpha = \alpha_c$, and for (b): $\theta_0 = 1.595$ for $\alpha < \alpha_c$, and $\theta_c = 0.396$ for $\alpha = \alpha_c$. The zero frequency peak $A \delta(\omega)$ of $C_{Sz}(\omega)$ is not shown. NRG parameters are $\Lambda=4.0$, $M_s=100$, $N_b=12$, and $B=1.0$ for broadening.
}   \label{Fig8}
\end{figure}

Focusing on the jointing point, we study in Fig.7 how the first-order QPT evolves into a continuous one as $\epsilon$ crosses $\epsilon_c$ from below. In Fig.7(a), $|\langle X \rangle|(\alpha)$ curves are shown for different $\epsilon$ values. As $\epsilon$ approaches $\epsilon_c$ from below, the jumps in $\langle X \rangle(\alpha_c^{(1)})$ (Fig.7(b)) and $\langle S_z \rangle(\alpha_c^{(1)})$ (Fig7.(c)) decreases to zero, first making a weak first-order QPT and then a continuous one. Note that the spin is always polarized on both sides of the QPT. Same as $\Delta=0$ case, in the limit $N_b=\infty$, $|\langle X \rangle| = \infty$ and $E_g = -\infty$ in the environment-unstable phase, regardless of the order of QPT. 

Besides the change of phase diagram, a finite $\Delta$ also induces non-trivial dynamics for $\sigma_z$ which is of utter importance for the realistic qubit experiments. The coherence in the non-equilibrium evolution $\langle \sigma_z \rangle(t)$ can be partly reflected in the equilibrium dynamical correlation function
\begin{equation}
   C_{Sz}(\omega)=\frac{1}{2\pi} \int_{-\infty}^{+\infty} C_{Sz}(t) e^{i\omega t} \, dt
\end{equation}   \label{eq18}
with $C_{Sz}(t) \equiv (1/2) \langle \{ S_z(t), S_z(0) \} \rangle$.\cite{Lv1} 
At $T=0$, $C_{Sz}(\omega) = C_{Sz}(-\omega)$ and it fulfils the sum rule $\int_{-\infty}^{\infty} C_{Sz}(\omega) d\omega = 1/4$. For a non-degenerate ground state $|G\rangle$, $C_{Sz}(\omega)= A \delta(\omega) + C^{\prime}_{Sz}(\omega)$, where $A= \langle G| Sz |G \rangle^{2}/2$. 
At $\Delta=0$, there is no dynamics in the $S_z$ component and $C_{Sz}(\omega)= \delta(\omega)/4$. For $\Delta > 0$, the spin is no longer fully polarized in $z$-direction and the weight of $C_{Sz}(\omega)$ is partially transferred from $\omega=0$ to $\omega > 0$ regime. 

In Fig.8(a) and (b), $C_{X}(\omega)$ and $C_{Sz}(\omega)$ are presented for $\epsilon = 0.0 > \epsilon_c$ and $\alpha \leqslant \alpha_c^{(c)}$. $C_{X}(\omega)$ shown in Fig.8(a) has the same low frequency asymptotic behaviour as $\Delta=0$, {\it i.e.}, $C_{X}(\omega) \propto \omega^{s}$ for $\alpha < \alpha_c^{(c)}$ and $C_{X}(\omega) \propto \omega^{-s}$ for $\alpha = \alpha_c^{(c)}$. $C_{Sz}(\omega)$ shown in Fig.8(b) has a high frequency peak, which represents the Rabi oscillation of a weakly damped qubit and it is not changed by $\alpha$ even at $\alpha_c$. In the low frequency regime, $C_{Sz}(\omega) \propto \omega^{\theta_0}$ for $\alpha < \alpha_c$ and $C_{Sz}(\omega) \propto \omega^{\theta_c}$ at $\alpha = \alpha_c^{(c)}$. NRG gives $\theta_0 = 1.595$ and $\theta_c = 0.396$ for $s=0.3$. For $\alpha < \alpha_c^{(c)}$, $C_{Sz}(\omega)$ has the same crossover scale $\omega^{\ast}$ as $C_{X}(\omega)$, separating the $\omega^{\theta_0}$ (for $\omega \ll \omega^{\ast}$) and $\omega^{\theta_c}$ (for $\omega \gg \omega^{\ast}$) behaviours. A zero frequency peak $A\delta(\omega)$ is also present (not shown here).

Close to the first-order QPT at $\epsilon < \epsilon_c$ and $\alpha \leqslant \alpha_c^{(1)}$, $C_{X}(\omega)$ and $C_{Sz}(\omega)$ are similar to the ones at $\epsilon > \epsilon_c$ and $\alpha < \alpha_c^{(c)}$. At $\alpha = \alpha_c^{(1)}$, both correlation functions change abruptly into an artefact of finite $N_b$. In the $\Delta=0$ case, the critical behaviour cannot be observed in $\epsilon < \epsilon_c$ and $\alpha < \alpha_c^{(1)}$ regime, because the lower subspace $\sigma_z=+1$ has no QPT. For $\Delta > 0$, due to the mixing of two subspaces, 
quantum critical behaviour can be observed in the intermediate frequency regime $\omega^{\ast} \ll \omega \ll \Omega_{R}$ for the weak first-order QPT at $\epsilon \lesssim \epsilon_{c}$ and $\alpha \lesssim \alpha_c^{(1)}$. 
The crossover scale $\omega^{\ast}$ decreases with increasing $\alpha$ and reaches a finite value at the first-order QPT $\alpha = \alpha_c^{(1)}$.
As $\epsilon$ approaches $\epsilon_c$ from below, $\omega^{\ast}(\alpha = \alpha_c^{(1)})$ decreases to zero and the first order QPT transits into continuous one.

\begin{figure}[t!]
\vspace*{-4.5cm}
\begin{center}
\includegraphics[width=5.9in, height=4.9in, angle=0]{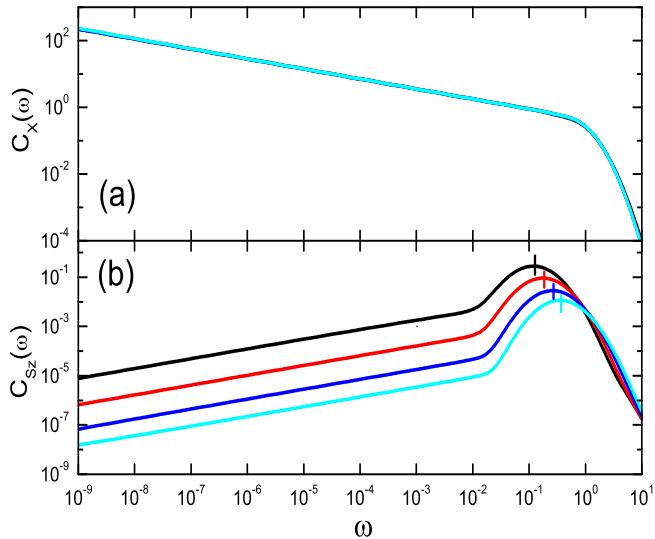}
\vspace*{-0.7cm}
\end{center}
\caption{(Color online) Dynamical correlation functions (a) $C_{X}(\omega)$ and (b) $C_{Sz}(\omega)$ for different $\epsilon > \epsilon_c \approx -0.098$, calculated for $s=0.3$, $\Delta=0.1$ and $\alpha=\alpha_c^{(c)}(\epsilon)$. From top to bottom, $\epsilon=-0.085$, $0.0$, $0.1$, and $0.2$. The corresponding $\langle S_z\rangle$ values are $-0.310$, $-0.421$, $-0.465$, and $-0.481$. The fitted exponents in small $\omega$ regime are $y_c = -0.298$ in (a) and $\theta_c = 0.396$ in (b). The zero frequency peak $A \delta(\omega)$ is not shown. In (b), the vertical dashes mark the Rabi frequency $\omega_R$ estimated from $\langle S_z\rangle$ and $\Delta_r \approx \Delta$. NRG parameters are $\Lambda=4.0$, $M_s=100$, $N_b=12$. The broadening parameter $B=1.0$.
}   \label{Fig9}
\end{figure}

Fig.9 shows the dynamical correlation functions $C_{X}(\omega)$ and $C_{Sz}(\omega)$ at the critical point $\alpha = \alpha_{c}^{(c)}$ for a series of $\epsilon$ in the regime $\epsilon > \epsilon_{c}$. Although $C_{X}(\omega)$ is independent of $\epsilon$, $C_{Sz}(\omega)$ decreases with increasing $\epsilon$, with the exponent unchanged. This is because as $\epsilon$ increases, $\langle S_z \rangle (\alpha = \alpha_{c}^{(c)})$ decrease monotonically to $-1/2$, transferring the weight of $C_{sz}(\omega)$ from the $\omega > 0$ regime to $\omega=0$. The prominent Rabi peak corresponds to short-time coherent oscillations in the population $P(t) = \langle S_z(t)\rangle$ of the non-equilibrium situation.~\cite{Lv1} The effective Rabi frequency $\omega_R$ increases with $\epsilon$. Assuming an effective free spin Hamiltonian $H_{eff}= (\epsilon_{eff}/2) \sigma_z - (\Delta_r/2) \sigma_x$, we can write $\omega_R = \sqrt{\epsilon_{eff} + \Delta_{r}}$ where $\epsilon_{eff}$ contains both $\epsilon$ and the static mean field from the quadratic coupling $(g_2/2) \sigma_z \langle Y^{2} \rangle$. $\Delta_r$ is the renormalized tunnelling strength. The estimated $\omega_R$ by assuming $\Delta_{r} \approx \Delta$ and using $\langle S_z \rangle$ from NRG agrees well with the peak position in $C_{Sz}(\omega)$ (vertical dashes in Fig.9). This shows that robust coherent spin evolution persists to the strongest coupling before the environmental QPT occurs.
The spin correlation function $C_{S_x}(\omega)$ for the dephasing properties of qubit was studied in Ref.~\onlinecite{Zheng2}. It was found that the high frequency peak in $C_{S_x}(\omega)$ has no change at $\alpha_c^{(c)}$, but broadened significantly only close to the spin flip line $\epsilon= \epsilon_f$, showing enhanced dephasing at the spin flip point.

\begin{figure}[t!]
\vspace*{-6.0cm}
\begin{center}
\includegraphics[width=6.8in, height=5.4in, angle=0]{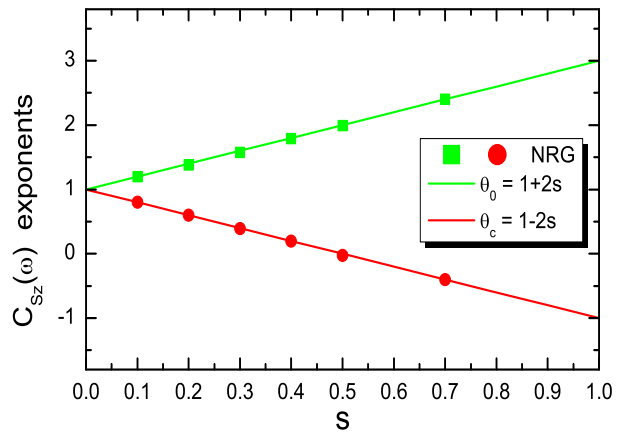}
\vspace*{-1.5cm}
\end{center}
\caption{(Color online) Exponents of $C_{S_z}(\omega)$: $\theta_0$ and $\theta_c$. The solid lines are $\theta_0=1+2s$ and $\theta_c=1-2s$.
}   \label{Fig10}
\end{figure}

\subsubsection{other $s$ values}

We carried out NRG study for other $s$ values and confirmed that the scenario of QPT established at $s=0.3$ applies to the whole sub-Ohmic regime $0 \leqslant s < 1$, with important quantitative differences. 

For $\Delta=0$, the structure of the phase diagram is same as that of $s=0.3$ and NRG results agree well with the exact solution. 
For $\Delta>0$, we find that the jointing point in the phase diagram moves upwards with increasing $s$. That is, $\epsilon_c$ increases with $s$ and for larger $s$, the first-order QPT line extends to larger $\epsilon$ values. At $T=0$, the critical fluctuation of $\hat{X}$ 
\begin{eqnarray}
  \langle X^{2} \rangle =   2 \int_{0}^{\infty} C_{X}(\omega) d\omega \propto \int_{0}^{\infty} \omega^{-s} d\omega 
\end{eqnarray}   \label{eq19}
increases with $s$. For larger $s$, the ground state energy contains a term $ \langle \sigma_z X^{2} \rangle$ which changes more rapidly with the flipping of spin. This makes the continuous QPT more difficult to realize. Our NRG study for $s=0.7$ supports that $\epsilon_c=\infty$ for any finite $\Delta$, {\it i.e.}, the transition is first-order for any $\Delta>0$ and any $\epsilon$, though this does not hamper the observation of power-law $C_{S_z}(\omega)$ in the intermediate frequency regime.~\cite{Zheng2} This behaviour is well understood in the extreme case $s \geqslant 1$ where the infra-red divergence in $\langle X^{2} \rangle$ makes the continuous QPT impossible. Here, the continuous versus first-order phase transition is an interesting problem on its own, giving its resemblance to the same problem in the crystal lattice.~\cite{Cowley1} A detailed study on this issue will be published elsewhere.
In the other limit $s=0$ which is related to the $1/f$ noise in the quantum circuit, a finite $\Delta$ induces a small but finite $\alpha_c^{(c)}$. $z\nu=1/s$ diverges at $s=0$ and a QPT of the Kosterlitz-Thouless type occurs, as confirmed by the NRG calculation (not shown). This is similar to the situation of linear-coupling SBM.~\cite{Bulla1}

The above analysis also explains the observation that for larger $s$, reliable NRG calculations require larger $N_b$ and are hence more difficult. For studying the continuous QPT at $\epsilon > \epsilon_c$, insufficient $N_b$ could lead to artificial critical fixed point and produce incorrect exponents $z\nu$, $\theta_0$ and $\theta_c$. For studying the first-order QPT at $\epsilon < \epsilon_c$, it may make an artificial continuous QPT. Up to now, quantitatively accurate study of $H_{QSB} (\Delta >0)$ for $s > 1$ is still a technical challenge for NRG. For the sub-Ohmic bath, however, we can get reliable results using the boson number truncation up to $N_b=50$ and a large logarithmic discretization parameter $\Lambda=10.0$. For $s \lesssim 1$ where the first-order QPT prevails, the critical exponents can still be extracted reliably from the intermediate frequency regime $\omega^{\ast} \ll \omega \ll \omega_{R}$ for $\epsilon \lesssim \epsilon_c$ and $\alpha \lesssim \alpha_c^{(1)}$ ({\it e.g.}, the data point for $s=0.7$ in Fig.10).

In Fig.10, we show the exponents $\theta_{0}$ and $\theta_c$ of $C_{Sz}(\omega)$. They are defined as $C_{Sz}(\omega) \propto \omega^{\theta_{0}}$ for $\alpha < \alpha_c^{(c)}$ and $C_{Sz}(\omega) \propto \omega^{\theta_{c}}$ for $\alpha = \alpha_c^{(c)}$. Since they appear only at $\Delta > 0$, there is no exact solution for them. The NRG data agree with the analytical expressions $\theta_{0}= 1-2s$ and $\theta_{c}=1+2s$ within an error of $2 \%$. This is in contrast to $\theta_0 = s$ and $\theta_c = -s$ for the linear-coupling SBM.~\cite{Bulla1,Vojta1} When extended to $s \geqslant 1$, such behaviour will lead to the breakdown of the sum rule of $C_{Sz}(\omega)$ and prohibit the continuous QPT in the Ohmic- and super-Ohmic regime.

\section{Discussion and Summary} 

In this section, we discuss several issues regarding to the impurity-induced environmental QPT that we studied in this paper. 

First, we note that the unphysical results $\langle X \rangle = \pm \infty$ and $E_g= -\infty$ in the boson-unstable state are the consequences of incompleteness of the present model. In reality, the boson number will not diverge even after the QPT occurs, because as the boson number increases, the interactions between boson modes that are neglected in our quadratic-coupling SBM, e.g., the anharmonic terms, will become important and finally keep the boson number from diverging. They will instead lead to a new stable state with finite $\langle X \rangle$, i.e., a state with local environmental distortion. Close to the environmental QPT on the weak-coupling side, the average boson number is small and these interactions play minor role. Therefore, the quadratic-coupling SBM Eq.(2) has a limited applicability range. It can be used to predict the existence of the impurity-induced QPT, to describe the phase diagram as well as the dissipation and dephasing effect due to the environmental fluctuation on the $\alpha < \alpha_c$ side, but cannot tell us what the exact ground state is in the parameter regime $\alpha > \alpha_c$.

 The environmental instability shows up differently in real systems. For the superconducting flux qubit system,~\cite{Bertet1} $\langle X\rangle \neq 0$ corresponds to an additional bias current in the SQUID oscillator. In the experiment of quantum dot system,~\cite{Muljarov1,Borri1} however, the boson instability corresponds to a local distortion of the crystal lattice. In the optical spectra signal of an impurity center in crystals, the instability is detected by the anomalous temperature dependence of the zero-phonon line width due to the softening of bosonic modes close to the environmental QPT.~\cite{Hizhnyakov1}
In the NRG calculation, the boson state truncation $N_b$ mimics such a higher order anharmonic effect accidentally. We find that although the existence of the QPT is robust under this constraint of Hilbert space, the critical exponents $z\nu$ and $\theta_c$ may well be changed by it.~\cite{Hou1,Tong2}

Second, we discuss the situation where both the linear- and the quadratic-coupling are present. In that case, the Hamiltonian reads
\begin{equation}
  H_{SB} = \frac{\epsilon}{2} \sigma_z -\frac{\Delta}{2} \sigma_x + \displaystyle\sum_{i} \omega_{i} a_{i}^{\dagger}a_{i} + \frac{g_1}{2}  \sigma_{z} \hat{Y} + \frac{g_2}{2} \sigma_{z} \hat{Y}^{2}.
\end{equation}   \label{eq20}
For general parameters $g_1 \neq 0$ and $g_2 \neq 0$, this Hamiltonian has a lower symmetry than both the linear-coupling SBM and the pure quadratic-coupling one. As a result, neither the delocalized-localized transition nor the environmental stable-unstable transition exists any more. Instead, similar to the situation of linear-coupling SBM under a finite bias $\epsilon$, it is expect that the ground state smoothly interpolates between different limiting symmetry-broken states of purely linear- or quadratic-coupling Hamiltonians. The crossover lines separating these phases are determined by the relative strength of $g_1$, $g_2$, and the crossover energy scale $T^{\ast}$ to the quantum critical regimes in $g_1$- and $g_2$-only cases.~\cite{Zheng1} However, both the bath and the spin dynamics will be severely influenced by the existence of the quadratic coupling terms.

At finite temperatures, the QPT observed in $H_{QSB}$ no longer exists, but turns into a crossover. At finite $T$, the quantum critical point at $T=0$ will expand into a finite parameter regime, the quantum critical regime, in which critical properties can be observed. The boundaries of this quantum
critical regime is determined by $T^{\ast}$, the crossover energy scale between the critical fixed point and the other stable fixed points. Temperature dependence of physical quantities will have the scaling form near the crossover. This scenario was verified in the linear-coupling SBM and was the basis for a proposal to observe the localied-delocalied QPT at finite temperature in a mesoscopic metal ring system.~\cite{Tong1} For the quadratic-coupling SBM studied in this work, we expect that the same scenario applies and can be used to observe the signature of the environmental QPT in experiment.

Our conclusion about the environmental QPT can be straightforwardly extended to the single boson mode case. For the Hamiltonian of the circuit quantum electrodynamics $H=\omega_p a^{\dagger}a + (\Delta/2) \sigma_x + (\epsilon/2) \sigma_z + g_2 (a^{\dagger} + a)^{2}\sigma_z$,~\cite{Bertet1} the boson-instability occurs at $g_2/\omega_p > 1/4$ for $\Delta=0$. Using the parameters of the experimental set up of Ref.~\onlinecite{Bertet1}, we estimate that $g_2 \sim 5.0$ MHz. Given $\omega_p= 3.17$GHz, the actual ratio $g_2/\omega_p \sim 10^{-3}$, much smaller the critical value. However, in the experiments of superconducting qubit, methods are available to engineer the shape and strength of $J(\omega)$ \cite{Haeberlein1} for a continuous environment, and to enhance the spin-boson coupling to the ultra-strong regime for discrete boson modes.~\cite{Niemczyk1} Especially, the new technique of switchable coupling can boost the linear coupling from $10^2$ MHz level to GHz level, making it comparable to $\omega_p$.~\cite{Peropadre1} The superconducting flux qubit~\cite{Yoshihara1,Kakuyanagi1} or the quantum dot~\cite{Petersson1} under the $1/f$ noise can also be tuned to the optimal working point . Considering that our results predict that the $1/f$ noise with quadratic spin-boson coupling gives a much smaller $\alpha_c$, we expect that these advances can make it feasible to detect the environmental QPT discussed in this work.

In summary, we predict a novel impurity-induced environmental QPT in the quadratic-coupling SBM which is realized in a wide class of experimental set ups. Using the exact solution at $\Delta=0$ as well as NRG, we obtain the ground state phase diagram which contains both continuous and first-order QPTs, with non-trivial critical properties. The dynamical correlation function of $\sigma_z$ is obtained, showing robust Rabi oscillation for $\alpha \leq \alpha_c^{(c)}$. Physical consequences of such QPTs and the feasibility of experimental observation are discussed.

\section{Acknowledgements}
D.-C. Zheng and N.-H. Tong acknowledge helpful discussions with Y.-J. Yan. 
This work is supported by 973 Program of China (2012CB921704), NSFC grant (11374362), Fundamental Research Funds for the Central Universities, and the Research Funds of Renmin University of China 15XNLQ03.

\appendix{}
\section{Exact solution at $\Delta=0$}

In this appendix, we derive the exact solution at $\Delta=0$. The Hamiltonian $H_{QSB}$ at $\Delta=0$ reads
\begin{equation}
    H_{QSB}(\Delta=0) =  \frac{\epsilon}{2} \sigma_z + \displaystyle\sum_{i} \omega_{i} a_{i}^{\dagger}a_{i} + \frac{g_2 \eta_0}{2\pi}  \sigma_{z} \hat{X}^{2}.
\end{equation}   \label{eq_A1}
Here $X= b_0 + b_0^{\dagger}$ and $b_0 = \sqrt{\pi/ \eta_0} \sum_{i} \lambda_i a_i$.  For the spectral function $J(\omega)$ specified in Eq.(4), $\eta_0 \equiv  \pi \sum_i \lambda_i^{2} =  2\pi\alpha \omega_c^{2}/(1+s)$.

To solve $H_{QSB}(\Delta=0)$ exactly, we employ the equation of motion (EOM) for the double-time Green's function $\langle \langle X|X\rangle \rangle_{\omega}$. It is defined as  $\langle \langle X|X\rangle \rangle_{\omega} \equiv \int_{-\infty}^{\infty} G^{r} \left[X(t)|X(t^{\prime}) \right] e^{i\omega t} dt$ and the retarded Green's function $G^{r} \left[ X(t)|X(t^{\prime}) \right] \equiv -i \theta (t) \langle [X(t), X(t^{\prime})]\rangle $. At zero temperature $T=0$, the dynamical correlation function $C_{X}(\omega)$ is expressed in terms of  $\langle \langle X|X\rangle \rangle_{\omega}$ as
\begin{equation}
   C_{X}(\omega) = - \frac{1}{2 \pi} {\text Sgn}(\omega)\text{Im}\langle \langle X|X \rangle \rangle_{\omega + i\eta} .
\end{equation}   \label{eq_A2}
Here $\eta$ is an infinitesimal positive number. $C_{X}(\omega)$ is an even function of $\omega$.

We start from the EOM of a GF component,
\begin{equation}
   \omega \langle \langle a_i|X\rangle \rangle_{\omega} = \langle \left[a_i, X \right] \rangle +  \langle \langle \left[ a_i, H_{QSB} \right] |X\rangle \rangle_{\omega}.
\end{equation}   \label{eq_A3}
At $\Delta=0$, the commutators in the above equation read $\left[a_i, X \right] = \sqrt{\pi/\eta_0} \lambda_i$ and
$\left[a_i, H_{QSB} \right] = \omega_i a_i + g_2 \lambda_i \sigma_z \sum_{l} \lambda_l \left(a_l + a_{l}^{\dagger} \right)$. Using these expressions and their Hermitian conjugates, we obtain
\begin{equation}
   \omega \langle \langle a_i + a_i^{\dagger} | X\rangle \rangle_{\omega} = \omega_i \langle \langle a_i - a_i^{\dagger} | X\rangle \rangle_{\omega}
\end{equation}   \label{eq_A4}
and 
\begin{eqnarray}
&&  \omega \langle \langle a_i - a_i^{\dagger} | X\rangle \rangle_{\omega} \nonumber \\
  &=& 2 \sqrt{\pi / \eta_0}\lambda_i + \omega_i \langle \langle a_i + a_i^{\dagger} | X\rangle \rangle_{\omega}   \nonumber \\
  && + 2 g_2 \sqrt{\eta_0/\pi}\sigma_z \lambda_i \langle \langle X | X \rangle \rangle_{\omega}.
\end{eqnarray}   \label{eq_A5}
One can solve Eq.(A4) and (A5) to obtain
\begin{eqnarray}
&&  \langle \langle a_i + a_i^{\dagger} | X\rangle \rangle_{\omega} \nonumber \\
&=& \sqrt{\frac{\pi}{\eta_0}} \frac{2\lambda_i \omega_i}{\omega^2 - \omega_i^{2}}  \nonumber \\
&& + g_2 \sqrt{ \frac{\eta_0}{\pi}}  \frac{2\lambda_i \omega_i}{\omega^2 - \omega_i^{2}} \sigma_z \langle \langle X | X \rangle \rangle_{\omega}.
\end{eqnarray}   \label{eq_A6}
Multiplying $\lambda_i$ on both sides of the above equation and summing over $i$, we obtain
\begin{equation}
   \langle \langle X | X \rangle \rangle_{\omega} = \frac{2\pi/\eta_0 \sum_i\frac{\lambda_i^{2} \omega_i}{\omega^2 - \omega_i^{2}} }{1- 2 g_2\sigma_z \sum_i \frac{\lambda_i^{2} \omega_i}{\omega^2 - \omega_i^{2}} }.
\end{equation}   \label{eq_A7}
Using $g(\omega)$ defined in Eq.(10) of the main text, we have
\begin{equation}
   \langle \langle X | X \rangle \rangle_{\omega} = \frac{ g(\omega) + g(-\omega) }{ 1- \frac{g_2 \eta_0}{ \pi} \sigma_z \left[ g(\omega) + g(-\omega) \right] }.
\end{equation}   \label{eq_A8}
Carrying out the analytical continuation $\omega \rightarrow \omega + i\eta$ and taking the imaginary part, we obtain the exact expression for $C_{X}(\omega)$ in Eq.(9) of the main text.
$g(\omega)$ can be simplified as
\begin{equation}
 g(\omega)  = \frac{1+s}{\omega_c} \mathcal{P}\int_{0}^{1} \frac{x^{s}}{\omega/ \omega_c - x} dx .
\end{equation}   \label{eq_A9}
Finally we obtain~\cite{Table}
\begin{equation}
   g(\omega) = \frac{1}{\omega} F(1, 1+s; 2+s; \frac{\omega_c}{\omega}).
\end{equation}   \label{eq_A10}
Here $F(\alpha, \beta; \gamma; z)$ is the hypergeometric function.
For the numerical calculation in $0< \omega < \omega_c$ and the analysis of $C_{X}(\omega=0)$, the above expression is transformed by analytical continuation into (for $\omega > 0$ ) 
\begin{eqnarray}
   g(\omega) &=& - \frac{1}{\omega_c} \frac{1+s}{s} F(1, -s; 1-s; \omega / \omega_c ) \nonumber \\
   && + \frac{\omega^{s}}{\omega_{c}^{1+s}} \cos\left[\pi (1+s) \right] \Gamma(2+s)\Gamma(-s),
\end{eqnarray}   \label{eq_A11}
and 
\begin{eqnarray}
   g(-\omega) &=& -\frac{1}{\omega_c} \frac{1+s}{s} F(1, -s; 1-s; -\omega / \omega_c )   \nonumber \\
   &&   - \frac{\omega^{s}}{\omega_{c}^{1+s}} \Gamma(2+s)\Gamma(-s).
\end{eqnarray}   \label{eq_A12}
Here $\Gamma(z)$ is the Gamma function. Series expansions can then be used for numerical evaluations, $F\left(1, -s; 1-s; z \right) = -\sum_{0}^{\infty} \left[s/(n-s) \right] z^n$ for $0 \leqslant z < 1$.~\cite{Table} Note that the analytical continuation does not apply to integer values of $s=0, 1, 2, ...$.

In order to calculate $E_g^{(+1)}$ and $E_g^{(-1)}$, we start from the expression at $\Delta = 0$
\begin{equation}
  E_g^{\sigma_z} = \frac{\epsilon}{2} \sigma_z + \frac{g_2 \eta_0}{2\pi}\sigma_z \langle X^2 \rangle + \sum_i \omega_i \langle a_i^{\dagger}a_i \rangle.
\end{equation}   \label{eq_A13}
The energy difference is 
\begin{eqnarray}
  \Delta E_g &\equiv & E_g^{(+1)} - E_g^{(-1)}     \nonumber \\
  &=& \epsilon + \frac{g_2 \eta_0}{2\pi} \left[ \langle X^2 \rangle^{(+1)} + \langle X^2 \rangle^{(-1)} \right]  \nonumber \\
  && + \sum_i \omega_i \left[ \langle a_i^{\dagger}a_i \rangle^{(+1)} - \langle a_i^{\dagger}a_i \rangle^{(-1)} \right].
\end{eqnarray}   \label{eq_A14}
The averages can be calculated from corresponding GFs using the fluctuation-dissipation theorem. For this purpose, besides $\langle \langle X | X \rangle \rangle_{\omega}$ obtained above, we still need $\langle \langle a_i | a_i^{\dagger} \rangle \rangle^{\sigma_z}_{\omega}$  which is obtained as
\begin{equation}
\langle \langle a_i | a_i^{\dagger} \rangle \rangle^{\sigma_z}_{\omega} = \frac{1}{\omega - \omega_i} + \frac{\lambda_i^2}{(\omega- \omega_i)^2 } \left[ g_2 \sigma_z + g_{2}^{2} \frac{\eta_0}{\pi}\langle \langle X|X \rangle \rangle_{\omega}  \right].
\end{equation}   \label{eq_A15}
The fluctuation-dissipation theorem at $T=0$ gives
\begin{equation}
   \Delta E_g = \epsilon + \frac{1}{\pi}
\int_{-\infty}^{0} \text{Im} H_{\omega + i\eta} d\omega,
\end{equation}   \label{eq_A16}
with
\begin{eqnarray}
   H_{\omega} &=&  \frac{g_2 \eta_0}{2\pi} \left[ \langle \langle X| X \rangle \rangle^{(+1)}_{\omega} + \langle \langle X|X \rangle \rangle^{(-1)}_{\omega} \right]  \nonumber \\
  && + \sum_i \omega_i \left[\langle  \langle a_i |  a_i^{\dagger} \rangle \rangle^{(+1)}_{\omega} - \langle \langle a_i | a_i^{\dagger} \rangle \rangle^{(-1)}_{\omega} \right].  \nonumber \\
  &&
\end{eqnarray}   \label{eq_A17}
Inserting Eq.(A8) and (A15) into this expression, we obtain the results Eqs.(14)-(16) of the main text.
Using the $J(\omega)$ in Eq.(4) and Taylor expanding Eq.(A16), we obtain
\begin{eqnarray}
   \Delta E_g & = & \epsilon + \frac{g_2}{\pi} \int_{-\omega_c}^{0} J(-\omega) d\omega + \mathcal{O}(\alpha^{3}) \nonumber \\
    &=& \epsilon + \frac{2 \alpha}{1+s} (g_2 \omega_c^{2})+ \mathcal{O}(\alpha^{3}).
\end{eqnarray}   \label{eq_A18}
The approximate spin-flip line Eq.(17) is obtained as the solution to $\Delta E_g = 0$.

\section{NRG Formalism for $H_{QSB}$}

In this appendix, we summarize the bosonic NRG formalism used to study $H_{QSB}$. Following the logarithmic discretization method of Bulla,~\cite{Bulla3} the Hamiltonian of the quadratic-coupling SBM can be mapped into the star-type Hamiltonian 
\begin{equation}
   H_{star} = \frac{\epsilon}{2} \sigma_z - \frac{\Delta}{2} \sigma_x + \frac{g_2}{2} \sigma_z \hat{Y^{2}} + \sum_{n=0}^{\infty} \xi_n a_n^{\dagger} a_n.
 \end{equation}   \label{eq_B1}
The local boson displacement operator $\hat{Y}$ is expressed as $\hat{Y} = \sum_{n=0}^{\infty} \left( \gamma_n/\sqrt{\pi} \right) \left(a_n^{\dagger} + a_n \right)$. The coefficients $\xi_n$ and $\gamma_n$ reads
\begin{equation}
  \xi_n = \frac{ \int_{\Lambda^{-(n+1)\omega_c}}^{\Lambda^{-n}\omega_c} \omega J(\omega) d\omega }{ \int_{\Lambda^{-(n+1)\omega_c}}^{\Lambda^{-n}\omega_c} J(\omega) d\omega },
\end{equation}   \label{eq_B2}
and 
\begin{equation}
   \gamma_n = \left[ \int_{\Lambda^{-(n+1)\omega_c}}^{\Lambda^{-n}\omega_c} J(\omega) d\omega  \right]^{1/2}.
\end{equation}   \label{eq_B3}
Carrying out an orthogonal transformation~\cite{Wilson1} for the boson modes, one obtains the Wilson-chain Hamiltonian
\begin{eqnarray}
   H_{chain} &=& \frac{\epsilon}{2} \sigma_z - \frac{\Delta}{2} \sigma_x + \frac{g_2}{2} \sigma_z \hat{Y}^{2}  \nonumber \\
   && + \sum_{n=0}^{\infty} \left[ t_n \left( b_n^{\dagger} b_{n+1} + b_{n+1}^{\dagger} b_{n} \right) + \epsilon_n b_{n}^{\dagger} b_n \right].   \nonumber \\
   &&  
\end{eqnarray}   \label{eq_B4}
Here $\hat{Y} = \sqrt{\eta_0/\pi} \left( b_0^{\dagger} + b_0 \right)$.
The coefficients $t_n$ and $\epsilon_n$ are expressed by the following recursive formula ($m \geqslant 0$)
\begin{equation}
    t_m = \left[ \sum_{n=0}^{+\infty} \left[ \left(\xi_n - \epsilon_m \right)u_{mn}  - t_{m-1}u_{m-1 n} \right]^{2}\right]^{1/2},
\end{equation}   \label{eq_B5}
\begin{equation}
    u_{m+1 n} = \frac{1}{t_m} \left[ \left(\xi_n - \epsilon_m \right)u_{mn}  - t_{m-1}u_{m-1 n} \right],
\end{equation}   \label{eq_B6}
and
\begin{equation}
   \epsilon_m = \sum_{n=0}^{+\infty} \xi_n u_{mn}^{2}.
\end{equation}   \label{eq_B7}
The initial condition for the recursive calculation is $t_{-1}=0$, $u_{-1 n}=0$, $u_{0n}= \gamma_n / \sqrt{\eta_0}$ with $\eta_0 = \sum_{n=0}^{+\infty} \gamma_{n}^{2}$.

The RG transformation is established for $N \geqslant 0$
\begin{eqnarray}
  && H_{N+1} =  \Lambda H_{N}  \nonumber \\
  &+& \Lambda^{N+1} \left[ t_N \left( b_{N}^{\dagger}b_{N+1} + b_{N+1}^{\dagger} b_N \right) + \epsilon_{N+1} b_{N+1}b_{N+1} \right], \nonumber \\
  &&  
\end{eqnarray}   \label{eq_B8}
with the starting Hamiltonian $H_0$ 
\begin{equation}
  H_0 = \frac{\epsilon}{2} \sigma_{z} - \frac{\Delta}{2} \sigma_x + \epsilon_0 b_{0}^{\dagger} b_0 + \frac{g_{2}}{2} \sigma_z \hat{Y}^2.
\end{equation}   \label{eq_B9}
The chain Hamiltonian is recovered in the limit of $N = \infty$ as $H_{chain} = \lim_{N \rightarrow \infty} \Lambda^{-N} H_{N}$.
The above NRG formalism is the same as that for the liner-coupling SBM, except that we replaced $\hat{Y}$ in the linear-coupling term with $\hat{Y}^{2}$.~\cite{Bulla1p}

\numberwithin{figure}{section}
\renewcommand{\thefigure}{C\arabic{figure}}

\section{Quantitative Comparison of NRG and exact solution at $\Delta=0$}

In this Appendix, we make quantitative comparison between NRG results and the exact solution at $\Delta=0$, for the order parameter $\langle X \rangle$, the phase boundaries, dynamical correlation function $C_{X}(\omega)$, and the exponents $z\nu$, $y_0$, and $y_{c}$. For this purpose, we extrapolate NRG data to the limit $\Lambda=1.0$, $M_s=\infty$ and $N_b=\infty$. 

\begin{figure}[t!]
\begin{center}
\includegraphics[width=5.4in, height=4.4in, angle=0]{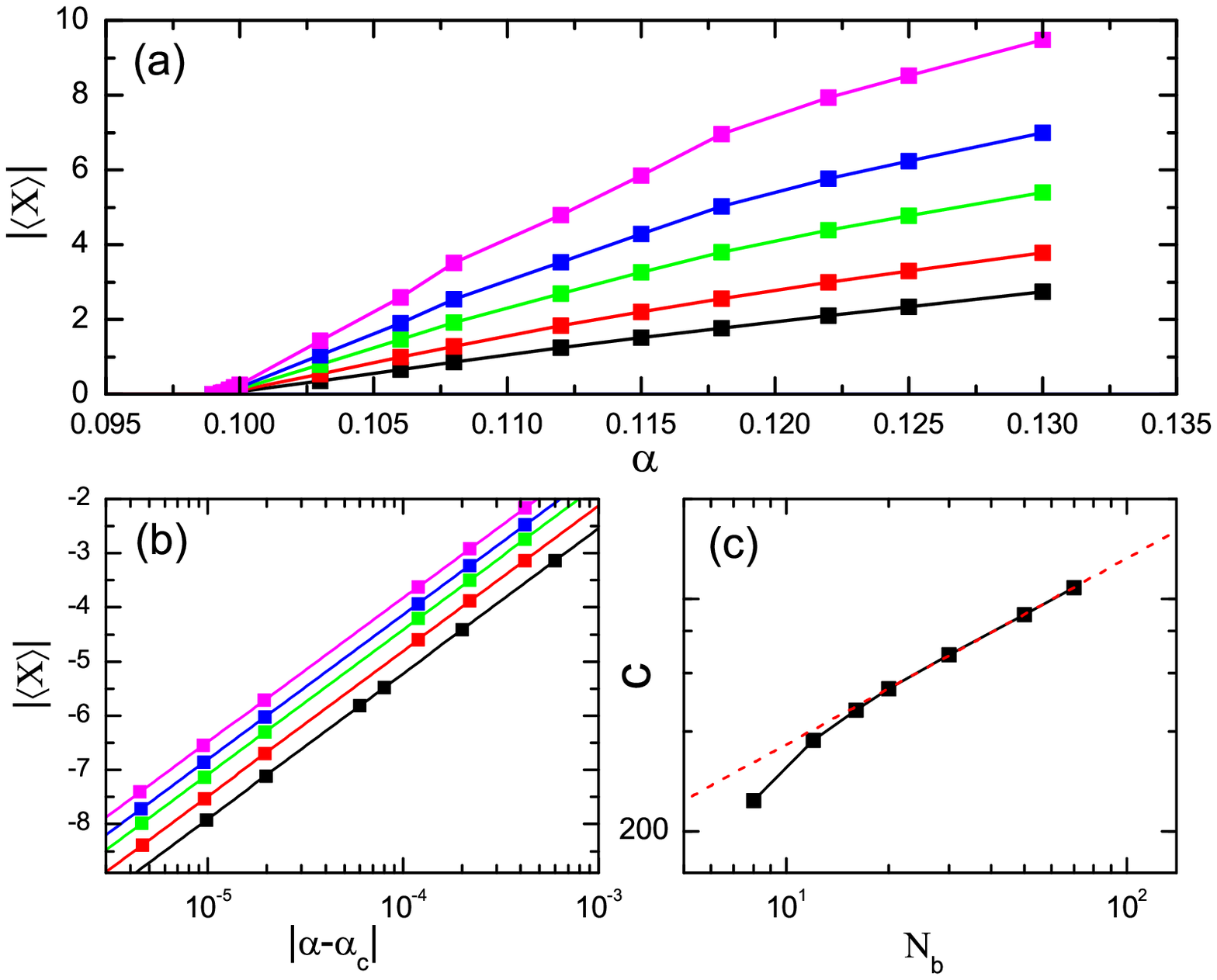}
\vspace*{-3.7cm}
\end{center}
\caption{(Color online) The critical behaviour of the order parameter $|\langle X \rangle|$ near the continuous QPT $\alpha = \alpha_c^{(c)}$, for $s =0.3$, $\Delta=0.0$, and $\epsilon=0.1 > \epsilon_c$. (a) $|\langle X \rangle|(\alpha)$ curves for different $N_b$ values. From bottom to top, $N_b=8, 12, 20, 30, 50$. (b) Power law fitting of $|\langle X\rangle| = c (\alpha-\alpha_c)^{\beta}$ with the fitted value $\beta = 1.16$ and $\alpha_c^{(c)}=0.09918$, being independent of $N_b$. (c) Log-log plot of the pre-factor $c$ versus $N_{b}$. The dashed line gives the fitting $c(N_b) \propto N_b^{0.56}$ in the large $N_b$ limit. The NRG parameters are $\Lambda=9.0$,  $M_s=100$.
}   \label{FigC1}
\end{figure}

%
\begin{figure}[t!]
\vspace*{-6.0cm}
\begin{center}
\includegraphics[width=6.1in, height=4.6in, angle=0]{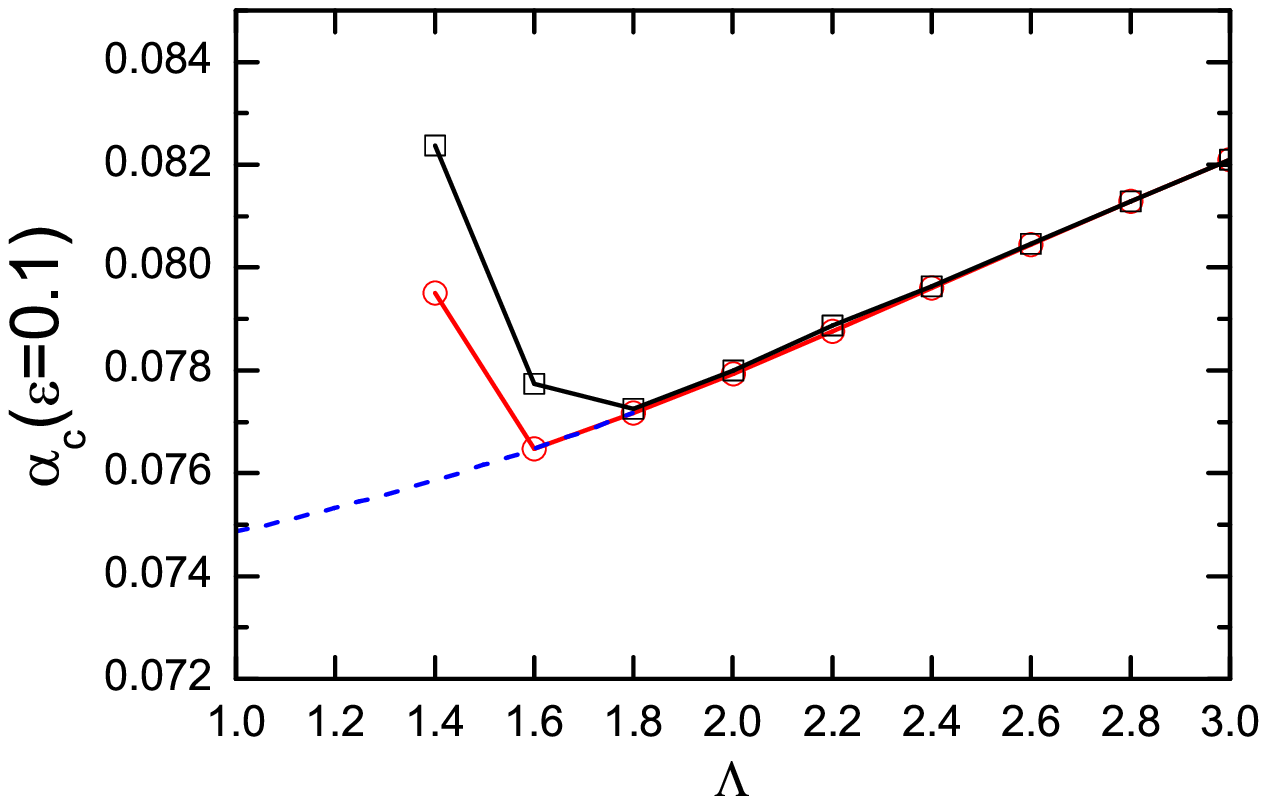}
\vspace*{-1.3cm}
\end{center}
\caption{(Color online) Extrapolation of $\alpha_c^{(c)}(\Lambda)$ to $\Lambda=1.0$ for $s=0.3$, $\Delta=0.0$, and $\epsilon=0.1 > \epsilon_c$. The squares and circles represent data obtained using $M_s=100$ and $Ms=200$, respectively, with $N_b=12$. The $4$-point Lagrangian extrapolation using data of $\Lambda=1.6$, $1.8$, $2.0$, and $2.2$ gives the dashed line and the extrapolated value $\alpha_c^{(c)}(\Lambda=1) = 0.0749$, in good agreement with the exact value $\alpha_c = s/(4g_2 \omega_c)=0.075$.
}
\end{figure}   \label{FigC2}

Fig.C1 shows the order parameter $\langle X \rangle$ and its critical behaviour near the continuous QPT at $\alpha =\alpha_c^{(c)}$ for $s=0.3$ and $\epsilon=0.1 > \epsilon_c$. Different $N_b$ values are used to extrapolate the results to $N_b = \infty$. We used a large discretization parameter $\Lambda=9.0$ so that the data are independent of $M_s$. As shown in Fig.C1(a), NRG always produces a finite $|\langle X \rangle|$ which increases with $N_b$. In Fig.C1(b), for each $N_b$ we show the critical behaviour $|\langle X \rangle| \propto c (\alpha - \alpha_c^{(c)})^{\beta}$ with the critical exponent $\beta=1.16$ and $\alpha_c^{(c)} = 0.09918$, both being $N_b$-independent. For general $0< s < 1$ we find that $\beta$ agrees well with $\beta = (1-s)/(2s)$, same as the expression due to boson state truncation obtained in the mean-field analysis of the linear-coupling SBM.~\cite{Hou1,Tong2} With increasing $N_b$, the pre-factor $c$ increases as $c(N_b) \propto N_{b}^{0.56}$ as shown in Fig.C1(c), leading to divergence of $|\langle X \rangle|$ at the critical point $\alpha = \alpha_{c}^{(c)}$ in the limit $N_{b}= \infty$.

\begin{figure}[t!]
\vspace*{-4.0cm}
\begin{center}
\includegraphics[width=5.3in, height=4.1in, angle=0]{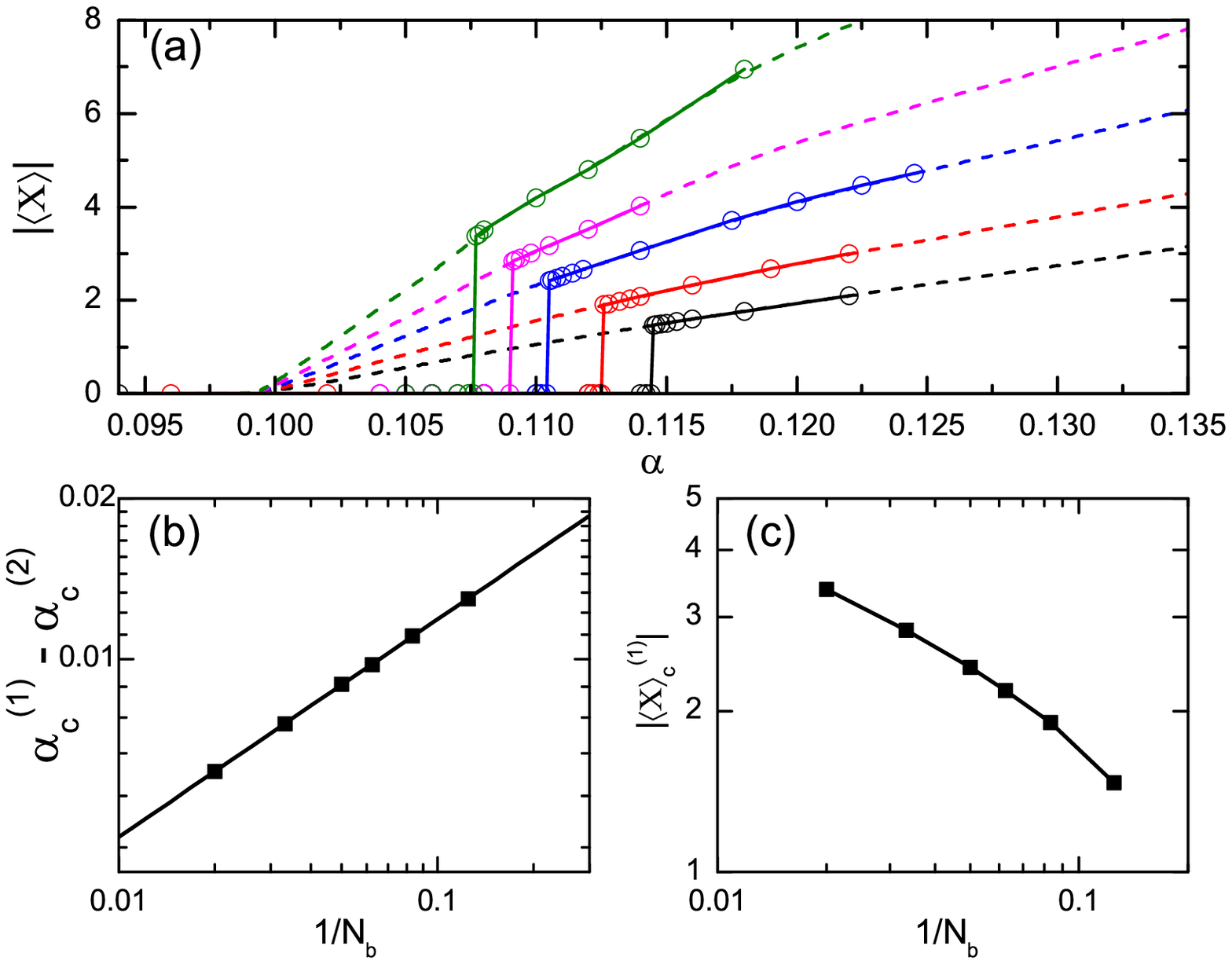}
\vspace*{-1.0cm}
\end{center}
\caption{(Color online) The curves $| \langle X \rangle|(\alpha)$ near the first-order QPT for $s =0.3$, $\Delta=0.0$, and $\epsilon=-0.2 < \epsilon_c$. 
In (a), the empty circles with eye-guiding lines are NRG data using various $N_b$'s. From bottom to top, $N_{b}=8, 12, 20, 30, 50$. The dashed lines are corresponding data for $\epsilon=0.1 > \epsilon_c$. (b) Distance of the first-order QPT point $\alpha_c^{(1)}$ to the continuous one $\alpha_c^{(c)}$ as a function of $1/N_b$. The solid squares are NRG data and the the solid line is a power law fit with the slope $-0.4$. (c) $|\langle X \rangle|$ value at the upper edge of $\alpha_c^{(1)}$ as a function of $1/N_b$. The NRG parameters are $\Lambda=9.0$ and $M_s=100$.
}   \label{FigC3}
\end{figure}

In Fig.C2, we extrapolate the NRG result $\alpha_c^{(c)}$ obtained for $\epsilon = 0.1 > \epsilon_c$ to the exact limit $\Lambda=1.0$. We plot the curves for $M_s=100$ and $M_s=200$, with a sufficiently large $N_b=12$. As shown in Fig.C2, a larger $M_s$ can produce reliable $\alpha_c^{(c)}$ down to smaller $\Lambda$ values. The $4$-point Lagrangian extrapolation of the $M_s=200$ data, using $\Lambda=1.6$, $1.8$, $2.0$, and $2.2$, gives $\alpha_c^{(c)}(\Lambda=1.0) = 0.0749$, very close to the exact result $\alpha_c^{exc} = s/(4g_2 \omega_c) = 0.075$ for $s=0.3$. For other $\epsilon > \epsilon_c$, the extrapolated values of $\alpha_c^{(c)}(\Lambda=1.0)$ coincide very well, being consistent with the $\epsilon$-independence of the exact $\alpha_c^{(c)}$.

The extrapolation of the first-order QPT point $\alpha_c^{(1)}$ to $N_b= \infty$ is demonstrated in Fig.C3 for $s=0.3$ and $\epsilon =-0.2 < \epsilon_c$. In Fig.C3(a), $|\langle X \rangle|(\alpha)$ curves (empty circles with eye-guiding lines) are plotted for different $N_{b}$ values. The same curves are also plotted for $\epsilon =0.1 > \epsilon_c$ (dashed lines) for comparison. For $\epsilon=-0.2$ and a fixed $N_b$, as $\alpha$ increases, $|\langle X \rangle|$ jumps at $\alpha = \alpha_c^{(1)}(N_{b})$ from zero to a finite value and then stays on the curve of $\epsilon=0.1$, for which a continuous QPT occurs at $\alpha_c^{(c)}=0.09918$. With increasing $N_b$, the non-zero $|\langle X \rangle|$ increases and $\alpha_c^{(1)}$ moves towards $\alpha_c^{(c)}$. Fig.C3(b) shows that the distance $|\alpha_c^{(1)}-\alpha_c^{(c)}| \propto N_{b}^{-0.4}$. This agrees with the conclusion from the exact solution, {\it i.e.}, the first-order QPT is a consequence of level-crossing made by the abrupt decrease of ground state energy in the $\sigma_z = -1$ subspace when a continuous QPT occurs and hence they have the same critical point $\alpha_c^{(1)} = \alpha_c^{(c)}$ in the limit $N_b = \infty$. $|\langle X \rangle_{c}^{(1)}|$ shown in Fig.C3(c) is the value at the upper edge of the the jump $\alpha = \alpha_{c}^{(1)}+ 0^{+}$. It also diverges in the limit $N_{b} = \infty$, as expected. All the NRG results up to now point to the conclusion that $|\langle X \rangle| = \infty$ once the environment enters the unstable state, irrespective of the order of QPTs.

\begin{figure}[t!]
\vspace*{-4.0cm}
\begin{center}
\includegraphics[width=5.7in, height=4.2in, angle=0]{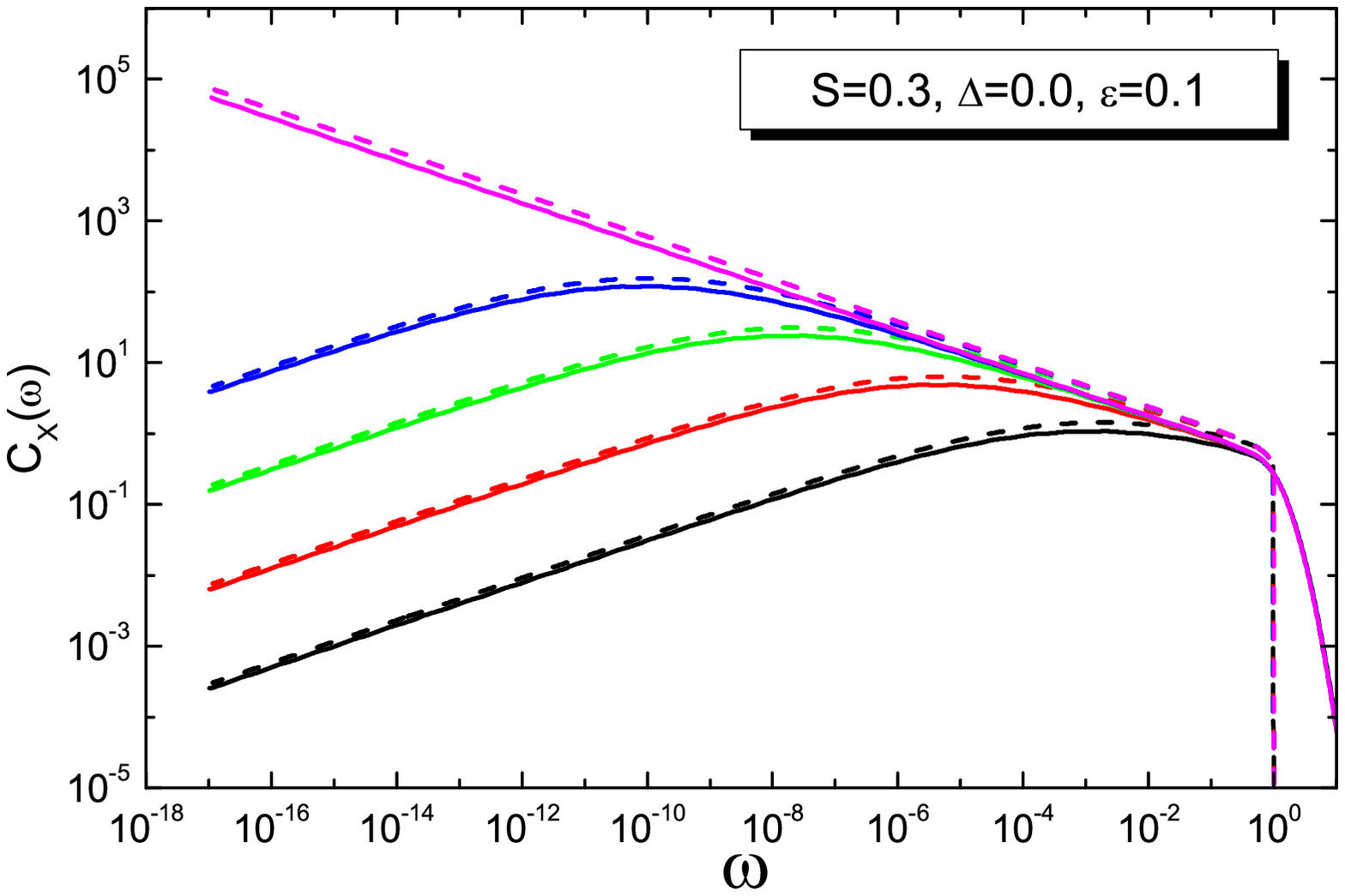}
\vspace*{-1.0cm}
\end{center}
\caption{(Color online) $C_{X}(\omega)$ for $s=0.3$, $\Delta=0.0$ and $\epsilon=0.1 > \epsilon_c$. The solid lines are NRG results and the dashed lines are exact solution Eq.(9) of the main text. From bottom to top, $\alpha = \alpha_c - \delta \alpha$ with $\delta \alpha = 1.0 \times 10^{-2}$, $2.0\times 10^{-3}$, $4.0 \times 10^{-4}$, $8.0 \times 10^{-5}$, and $0.0$. For NRG and the exact solution, we use respectively $\alpha_c^{NRG} = 0.085934484$ and $\alpha_c^{exc}=0.075$. The NRG parameters are $\Lambda=4.0$, $M_s=100$, $N_b=12$. The broadening parameter is $B=1.0$.
}   \label{FigC4}
\end{figure}

\begin{figure}[t!]
\vspace*{-5.0cm}
\begin{center}
\includegraphics[width=6.1in, height=5.0in, angle=0]{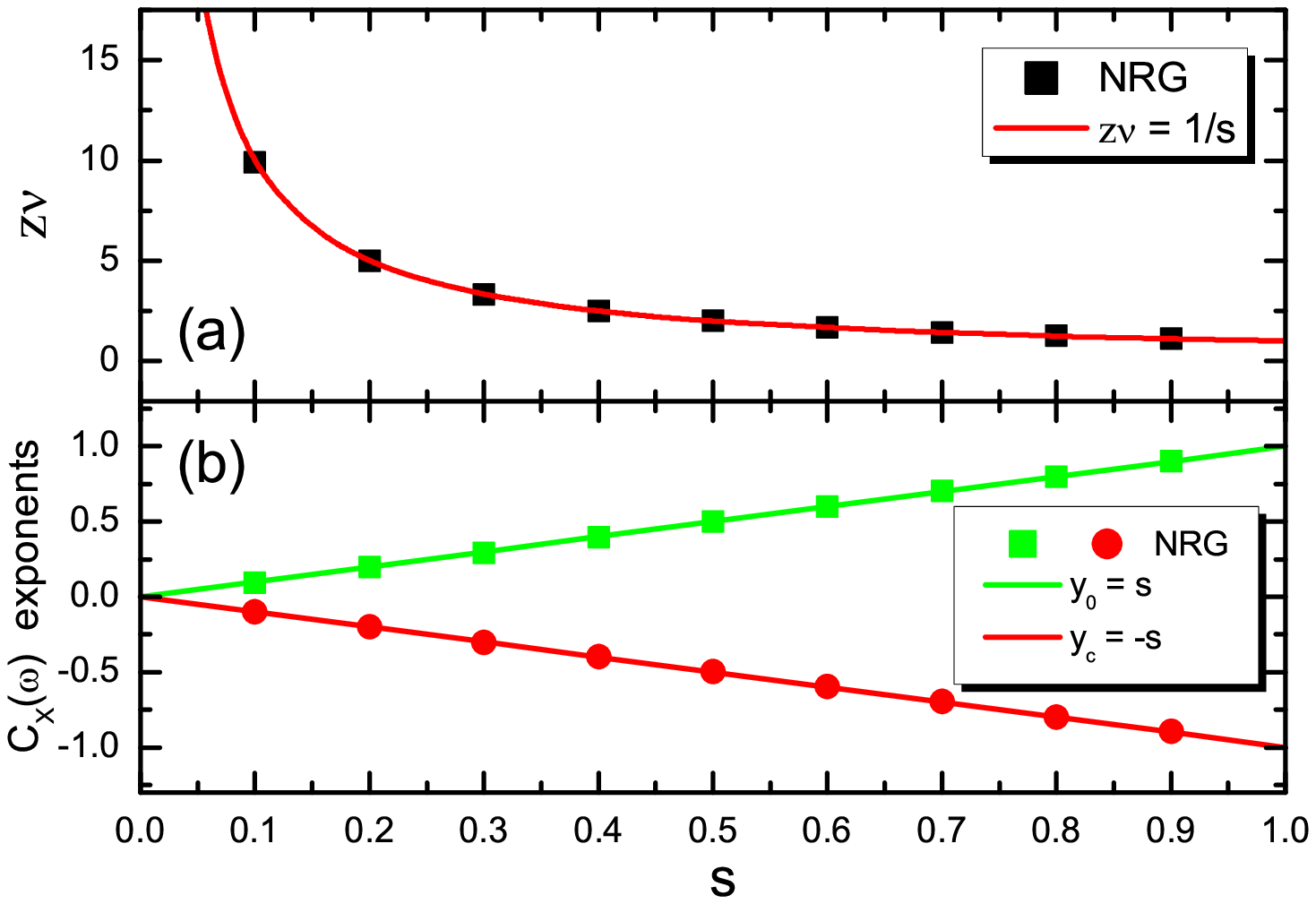}
\vspace*{-1.0cm}
\end{center}
\caption{(Color online) Comparison of the universal exponents calculated from NRG and the exact solution. (a) The energy scale exponent $z\nu$. Symbols are NRG data and the solid line $z\nu=1/s$ is the exact solution; (b) exponents of $C_{X}(\omega)$: $y_0$ and $y_c$.  Symbols are NRG data and the solid lines are exact solution $y_0 = s$ and $y_c= -s$.
}   \label{FigC5}
\end{figure}

In Fig.C4, we compare the NRG result for the dynamical correlation function $C_{X}(\omega)$ with the exact one obtained from Eq.(9) in the main text. Calculated at the same distance to the respective critical point $\alpha_c^{NRG}$ and $\alpha_c^{exc}$, the NRG results obtained using $\Lambda=4.0$, $M_s=100$, and $N_b=12$ and the exact solution agree quite well in the power law, the scaling form, and the crossover frequency. Quantitatively, the NRG results are smaller uniformly by $30\%$ in magnitude. This error comes mainly from the discretization error and can be reduced by extrapolating $\Lambda$ to unity. In the low frequency limit, the power law behaviour $\omega^{s}$ for $\omega \ll \omega^{\ast}$ and $\omega^{-s}$ for $\omega \gg \omega^{\ast}$ are clearly seen, with a crossover scale $\omega^{\ast}$ approaching zero as $\alpha$ tends to $\alpha_c^{(c)}$. The exact $C_{X}(\omega)$ curve has a sharp cut-off at $\omega=\omega_c$, inherited from the hard cut-off of $J(\omega)$ in Eq.(4). The long tail of the NRG curves in $\omega > 1.0$ is an artefact from the log-Gaussian broadening used in NRG.

In Fig.C5, we compare the critical exponents obtained from NRG (solid symbols) with the exact expressions (solid lines) in the range $0<s<1$. $z\nu$ shown in Fig.C5(a) is the critical exponent of the crossover energy scale $T^{\ast} \propto |\alpha - \alpha_c^{(c)}|^{z\nu}$. The NRG data agree well with the exact expression $z\nu=1/s$. In Fig.C5(b), the NRG results for $y_{0}$ and $y_{c}$ are compared with the exact expressions $y_{0}=s$ and $y_{c}=-s$. Here $y_{0}$ and $y_{c}$ are the low frequency exponent of $C_{X}(\omega)$: $C_{X}(\omega) \propto \omega^{y_{0}}$ for $\alpha < \alpha_c^{(c)}$ and $C_{X}(\omega) \propto \omega^{y_{c}}$ at the critical point $\alpha = \alpha_c^{(c)}$.

In summary, in this Appendix we made detailed comparison between NRG and the exact solution for $\Delta=0$ and good quantitative agreement is achieved.

\end{document}